\documentclass{article}

\usepackage{makecell}
\usepackage{ulem}
\usepackage{booktabs}
\usepackage{natbib}
\usepackage{graphicx}
\usepackage{fullpage}
\usepackage{multicol}
\usepackage{hyperref}
\usepackage{booktabs}
\usepackage{subcaption} 
\usepackage{float} 
\usepackage{makecell}
\usepackage{array}
\usepackage{threeparttable}
\usepackage{amssymb}
\usepackage{amsmath}
\usepackage{multirow}

\usepackage[dvipsnames,table,xcdraw]{xcolor}

\begin{document}

\title{Selecting ChIP-seq Normalization Methods from the Perspective of their Technical Conditions}

\author{Sara Colando, Danae Schulz, Johanna Hardin}

\abstract{Chromatin immunoprecipitation with high-throughput sequencing (ChIP-seq) provides insights into both the genomic location occupied by the protein of interest and the difference in DNA occupancy between experimental states. Given that ChIP-seq data is collected experimentally, an important step for determining regions with differential DNA occupancy between states is between-sample normalization. While between-sample normalization is crucial for downstream differential binding analysis, the technical conditions underlying between-sample normalization methods have yet to be examined for ChIP-seq. We identify three important technical conditions underlying ChIP-seq between-sample normalization methods: balanced differential DNA occupancy, equal total DNA occupancy, and equal background binding across states. To illustrate the importance of satisfying the selected normalization method's technical conditions for downstream differential binding analysis, we simulate ChIP-seq read count data where different combinations of the technical conditions are violated. We then externally verify our simulation results using experimental data. Based on our findings, we suggest that researchers use their understanding of the ChIP-seq experiment at hand to guide their choice of between-sample normalization method. Alternatively, researchers can use a high-confidence peakset, which is the intersection of the differentially bound peaksets obtained from using different between-sample normalization methods. In our two experimental analyses, roughly half of the called peaks were called as differentially bound for every normalization method. High-confidence peaks are less sensitive to one's choice of between-sample normalization method, and thus could be a more robust basis for identifying genomic regions with differential DNA occupancy between experimental states when there is uncertainty about which technical conditions are satisfied.}

\maketitle

\section{Introduction}

In recent decades, high-throughput sequencing has become one of the most popular methods for data generation in genomics, epigenomics, and transcriptomics \citep{overview-high-throughput}. A popular method of high-throughput sequencing is chromatin immunoprecipitation with high-throughput sequencing (ChIP-seq). ChIP-seq typically involves shearing the DNA via sonication before conducting immunoprecipitation with an antibody that is known \textit{a priori} to bind with the protein of interest (i.e., transcription factor, histone mark, etc.). Ideally, only the DNA fragments that are occupied by the protein of interest will remain after immunoprecipitation. These remaining fragments are then purified and sequenced before being aligned to a reference genome \citep{data-collection}. The aligned \textit{reads} are then used to characterize the amount of occupancy of the protein of interest within a specific genomic region \citep{Wu}. ChIP-seq experiments can be used as a binary measure of whether the protein of interest is bound or unbound at a genomic location. However, we focus on a different goal of many ChIP-seq experiments: comparing the amount of binding (at a particular site) across two experimental states (differential binding analysis). While identifying binding sites is the most common goal of ChIP-seq experiments, many researchers investigate questions about differential binding using ChIP-seq data (for example, see \cite{bornelov2018,chen2019,xiaoyang2024, shang-kun2022,holmes2022,kumar2023,mohammed2015,nagarajan2020,ross-innes2012,sabbagh2018,shah2021,siersbaek2020}).

Given that ChIP-seq experiments are conducted to assess which genomic regions are truly differentially occupied by the protein of interest, regions enriched with DNA binding, peaks, typically serve as the unit of interest in ChIP-seq data analysis. In this paper, we refer to \textit{DNA occupancy (per cell)} as the population parameter that we aim to estimate using ChIP-seq analysis and the sample estimate of the DNA occupancy (per cell) as the \textit{DNA binding (per cell)}. In this sense, a peak is considered \textit{differentially bound} if there is a statistically significant difference in the amount of DNA binding in the peak region between the experimental states \citep{workflow}.\footnote{In current literature, the population parameter and its sample estimate are usually both referred to as `DNA binding (per cell)'. However, we use distinct terms in this paper to disambiguate when we are referring to the population parameter (DNA occupancy) versus its sample estimate (DNA binding).} The raw amount of DNA binding within a given peak is calculated by counting the number of reads aligned to the specific genomic region. On average, genomic regions with higher read counts, i.e., more DNA binding (per cell), are expected to have a higher amount of DNA occupancy (per cell) \citep{workflow}.

That said, since ChIP-seq data is collected experimentally, there are expected to be differences in the observed DNA binding between the experimental states, even if there is the same amount of DNA occupancy in the states. Thus, hypothesis tests are essential for determining whether there is sufficient evidence that a difference in DNA binding between experimental states reflects a true difference in DNA occupancy between the states. The process of performing hypothesis tests to identify statistically significant differences in the DNA binding between experimental states is called \textit{differential binding analysis} and is popular for analyzing other types of high-throughput data beyond ChIP-seq. For instance, Cleavage Under Targets and Release Using Nuclease (CUT\&RUN) also leverages differential binding analysis to assess changes in protein-DNA interaction between experimental states \cite{workflow, skene2017}.\footnote{CUT\&RUN produces very similar data to ChIP-seq. The difference between the techniques is that in CUT\&RUN, DNA fragments are generated by adding an anti-IgG-micrococcal nuclease fusion protein after the primary antibody is added to the target of interest, rather than by sonicating the DNA as in ChIP-seq. After binding to the primary antibody, the fusion protein cuts the DNA around the target protein, producing sequenced fragments in CUT\&RUN. CUT\&TAG is a variation that utilizes tagmentation to generate cut fragments rather than using micrococcal nuclease.}

Performing accurate differential binding analysis requires the raw read counts in each peak to be normalized between samples, since the raw read counts can be affected by experimental artifacts, such as variations in the amount of DNA loaded or the quality of the antibody used between samples \citep{Peak_Call}. Such experimental artifacts have the potential to influence the sequencing (or read) depth (i.e., the total number of reads in the entire sample) for the entire sample \citep{workflow}, meaning that differences in the raw read counts for a given peak can arise between experimental states even when there is no difference in the DNA occupancy between the states. (Note that other normalization methods, for example those that address GC content bias, should also be used when the analysis goals are not focused on comparing DNA occupancy of a single region across different experimental conditions \cite{teng2017}, but in our work we focus only on between-sample normalization methods.)

There are various between-sample ChIP-seq normalization methods available to researchers (e.g., spike-in methods, background-bin methods, and peak-based methods \cite{DiffBind-vignette}). While the role of normalization methods has been analyzed through the lens of their technical conditions for RNA-seq \cite{Ciaran}, another popular type of high-throughput sequencing, there has been no parallel analysis of the between-sample normalization methods for ChIP-seq. However, ChIP-seq and RNA-seq data differ from each other in crucial ways. For one, RNA-seq focuses on characterizing (differential) gene expression, while ChIP-seq focuses on characterizing (differential) DNA occupancy. As a result, ChIP-seq data does not have predefined genomic regions of interest, whereas genes serve as the genomic regions of interest in RNA-seq data. Indeed, the genomic regions of interest in ChIP-seq are defined through the process of \textit{peak calling}, which leverages hypothesis testing (or other statistical tools, such as Hidden Markov Models \cite{allhoff2014}) to identify genomic regions that are significantly enriched with DNA binding \cite{MACS}. Further, because the experimental processing of ChIP-seq samples involves many steps over multiple days, and because both antibody quality and cell number contribute to the level of background noise, the signal-to-noise ratio can be quite variable between samples in ChIP-seq data compared to that in RNA-seq data \cite{Kidder2011}. For these reasons, we cannot directly apply the results from previous RNA-seq between-sample normalization analyses to ChIP-seq between-sample normalization methods. Rather, between-sample normalization methods available for ChIP-seq data and other similarly structured types of high-throughput data must be explicitly analyzed through the lens of their own technical conditions.

In this paper, we identify three key technical conditions underlying between-sample normalization methods for ChIP-seq and similarly structured types of high-throughput data: {\bf (1)} balanced differential DNA occupancy, {\bf (2)} equal total DNA occupancy across experimental states, {\bf (3)} equal background binding across experimental states. We simulate ChIP-seq read count data to demonstrate how violating these technical conditions can substantially impact the accuracy of downstream differential binding, leading to differential binding analysis with higher empirical false discovery rates and lower power, on average. We also use experimental CUT\&RUN \cite{ashby2022} and ChIP-seq \cite{ross-innes2012} data to validate that normalization methods which rely on the same technical condition yield similar normalization results and demonstrate the applicability of using a high-confidence peakset that only contains the peaks would be identified as differentially bound regardless of the between-sample normalization method used. Based on our results, we recommend that researchers use their understanding of the experiment at hand to guide their choice of ChIP-seq between-sample normalization method when possible. Alternatively, researchers could conduct analogous data analyses in which they only vary the selected between-sample normalization method when there is uncertainty about which technical conditions are violated. A separate differentially bound peakset would then be generated for each between-sample normalization method. Then, the researcher could use an intersection of these peaksets to create a high-confidence peakset which would be more robust to violations of the between-sample normalization technical conditions than the differentially bound peakset associated with any singular between-sample normalization method. This, in turn, would limit the impact of the choice of between-sample normalization method on the biological conclusions drawn from downstream differential binding analysis.

\section{Normalization and Differential Binding Analysis}

In a typical ChIP-seq workflow, samples are cross-linked with formaldehyde to fix DNA-bound proteins, and the chromatin is sheared. Antibodies are added to the protein of interest to immunoprecipitate bound DNA, which is then separated from the proteins and sequenced. Input DNA is collected prior to immunoprecipitation as a control. Additionally, some researchers will use an irrelevant IgG antibody for immunoprecipitation as an extra control. After the DNA is sequenced, it is aligned to the reference genome and analyzed using a peak caller. Peak callers typically compare the amount of DNA immunoprecipitated with the antibody against the protein of interest to the DNA collected from the input control or the IgG control antibody to call regions of the genome with statistically significant pile-ups of reads, or peaks for the immunoprecipitated DNA \cite{Kidder2011}. These regions with statistically significant DNA binding are assumed to be bound by the protein of interest. Additional controls with knockout strains can help to remove any spurious peaks from the dataset. Once the peaks are identified for a given experimental state, researchers might wish to interrogate whether the DNA occupancy for the protein of interest varies between experimental states \cite{bornelov2018,chen2019,xiaoyang2024, shang-kun2022,holmes2022,kumar2023,mohammed2015,nagarajan2020,ross-innes2012,sabbagh2018,shah2021,siersbaek2020}. 

In ChIP-seq differential binding data analysis, between-sample normalization is performed on the \textit{consensus peak set across experimental states}; that is, the subset of peaks that are consensus peaks in at least a certain number of experimental states. For a peak to be considered a \textit{consensus peak within an experimental state}, it must be called as a peak within at least a certain number (or proportion) of \textit{replicates} \cite{Landt2012}. After the consensus peaks across experimental states have been identified, a read count matrix can be generated, where each entry corresponds to the number of reads aligned to a given consensus peak in a given sample \citep{DiffBind-vignette}. The raw read counts are then normalized between samples and, sometimes also within samples. For between-sample normalization, the normalized read count for consensus peak $i$ in sample $j$ is typically computed as follows, where $s_j$ denotes the sample-wide size factor that corresponds to sample $j$:\footnote{Some methods that we consider in this paper, such as MAnorm2 (Common Peaks) and Loess Adjusted Fit (Reads in Peaks), normalize the raw reads by scaling them via (local) regression instead of generating a sample-wide size factor for each replicate.}

\begin{equation}
    k^*_{ij} = \frac{k_{ij}}{s_j}
\label{eq:norm_factor_general}
\end{equation}

The overarching goal of between-sample normalization for ChIP-seq is to eliminate discrepancies in the read counts between sample groups that are due to experimental artifacts rather than genuine changes in the DNA occupancy between experimental states. Experimental artifacts, such as the amount of DNA loaded into the sequencer, the quality of the antibody, the starting cell number, and how many reads are returned from the sequencer, can influence how many reads are aligned to a particular genomic region within a given sample \citep{workflow}. As a result, between-sample normalization is considered to have correctly normalized the ChIP-seq data between experimental states when the relationship between the samples' normalized read counts tracks the true relationship between the states' DNA occupancy levels. In other words, between-sample normalization is considered accurate if peaks that do not have differential DNA occupancy between samples should have the same normalized read counts across experimental states, on average, and peaks that do exhibit differential DNA occupancy between states have different normalized read counts across experimental states, on average. Figure~\ref{fig:oneOverExpressedPeak} demonstrates the difference between incorrect and correct ChIP-seq between-sample normalization through a toy example, where we assume peak calling has already been performed (e.g., using input or IgG controls) (figure adapted from Evans et al.\ \cite{Ciaran}).

\subsection{Importance of Correct Normalization to Differential Binding Analysis}

\begin{figure}[H]
    \centering
    \includegraphics[width=\linewidth, alt={A diagram with five subplots that walks through a toy example which shows the importance of correct between-sample normalization to downstream differential binding analysis. In all of the subplots, the color corresponds to the peak number (1, 2, or 3). Subplot (a) is a bar plot of the DNA occupancy per cell split into experimental states A and B, with bars corresponding to peak number. Subplot (b) is a proportional bar plot of the share of DNA binding across the three peaks in state A versus state B. Subplot (c) is a synthetic diagram of the reads aligned to each peak, where each rectangular tick mark represents a single read aligned to the peak region in the particular experimental state. Subplot (d) is split into three bar plots: no normalization, Library Size (Reads in Peaks) normalization, and correct normalization. In each of the bar plots in subplot (d), the horizontal axis is the peak and given experimental state, and the vertical axis is the normalized read count. Subplot (e) is a bar plot of the fold change (computed as normalized reads in state A divided by normalized reads in state B) for the no normalization, Library Size (Reads in Peaks) normalization, and correct normalization.}]{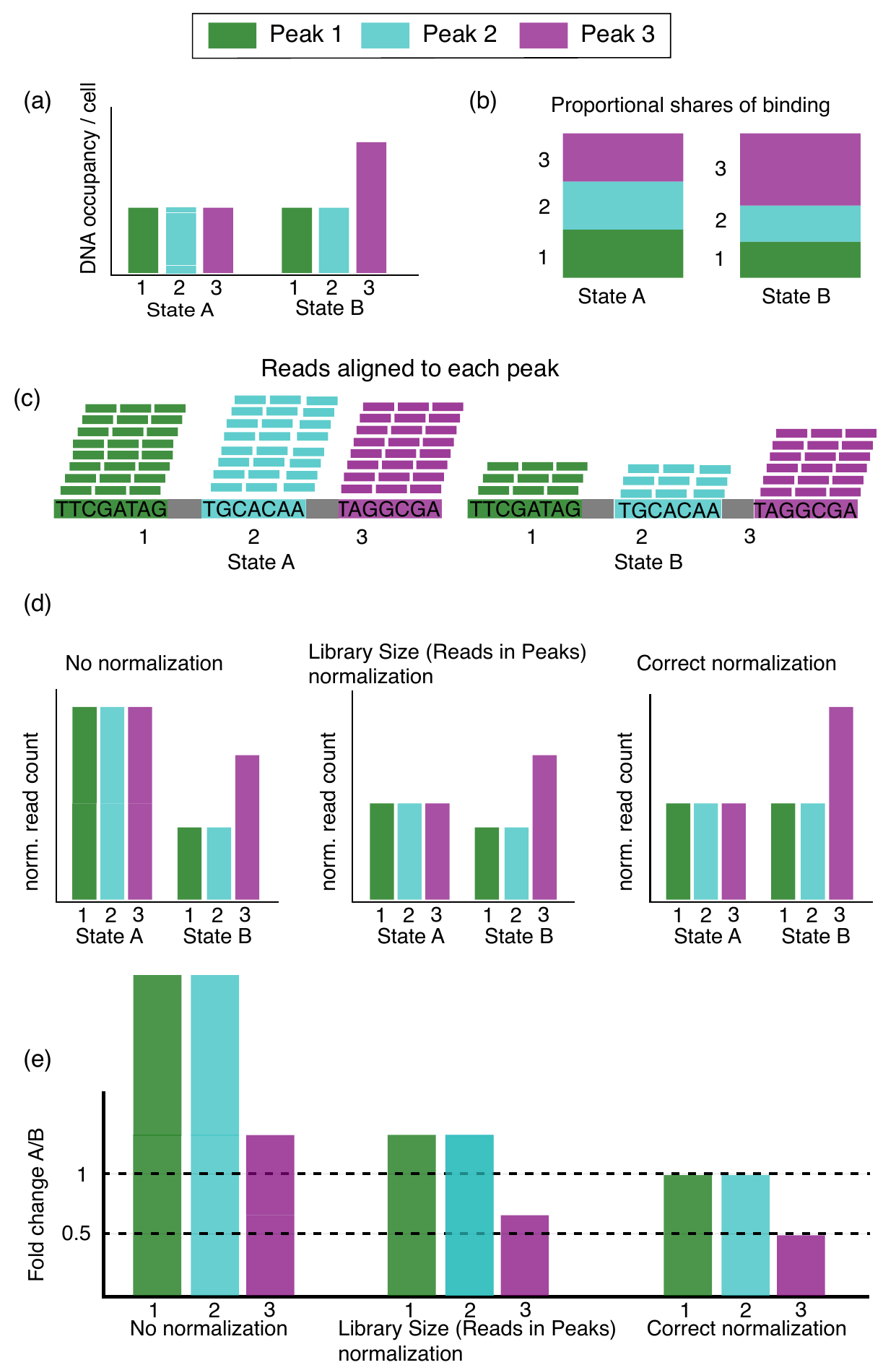}
    \caption{A toy example showing the importance of correct normalization to downstream differential binding analysis. {\bf(a)} depicts the true DNA occupancy per cell for the three peaks across experimental states A and B. Only Peak 3 (in purple) has differential DNA occupancy between A and B, with a higher amount of DNA occupancy per cell in state B. However, states A and B have an unequal amount of total DNA occupancy per cell. {\bf (b)} describes the proportional shares of DNA occupancy in the three peaks in both experimental states, {\bf (c)} depicts the reads aligned to each peak in each state, and {\bf (d)} depicts the normalized read count associated with each peak: the states for no normalization (i.e., the normalized reads are simply the raw reads), Library Size (Reads in Peaks) normalization, and correct normalization. {\bf (e)} shows the fold change A/B associated with each peak for the various normalization methods. Comparing {\bf (a)} and {\bf (c)}, we see that the total library size is the same across states A and B, even though state B has more DNA occupancy per cell. Correct normalization returns the true fold change A/B relationship in the DNA occupancy for all three peaks, i.e., the fold change A/B relationship in {\bf (a)}. Library Size (Reads in Peaks) normalization does not return the true fold change relationship in {\bf (a)} and indicates that there is a change in DNA occupancy per cell in all three peaks. Figure adapted from Evans et al.\ \cite{Ciaran}.}
    \label{fig:oneOverExpressedPeak}
\end{figure}

Correct between-sample normalization is crucial for meaningful differential binding analysis. Indeed, Wu et al.\ note that out of all the ChIP-seq data analysis steps they analyzed, between-sample normalization has the greatest potential to influence the discovery of differentially bound regions \cite{Wu}. (Relatedly, Reske et al.\ demonstrate that normalization method can significantly affect differential accessibility analysis and interpretation in ATAC-seq data \cite{reske2020}.) Our focus in this paper is to interrogate common between-sample normalization methods after peaks have already been called using appropriate controls. One important experimental method for normalizing between ChIP-seq samples is through the addition of spike-in DNA controls, where an equal amount of DNA that does not correspond to the genome of interest is added to each experimental sample prior to the immunoprecipitation step, which can help control for differences in genomic shearing or immunoprecipitation efficiency \cite{Orlando2014}. However, spike-in DNA is not always included for every ChIP-seq experimental dataset. Figure~\ref{fig:oneOverExpressedPeak} walks through a toy example with no spike-in data, where a standard normalization technique, Library Size (Reads in Peaks), leads to incorrect between-sample normalization, and consequently incorrect differential binding analysis results. In the toy example, there is unequal  DNA occupancy between experimental states A and B, with state B having a higher amount of DNA occupancy than state A. Additionally, only Peak 3 exhibits differential DNA occupancy between the experimental states, with it having double the amount of DNA occupancy in state B compared to in state A. As we highlight later on, Library Size (Reads in Peaks) relies on the technical condition that there is equal DNA occupancy across the experimental states. As a result, when the read counts in the toy example are normalized via Library Size (Reads in Peaks), all three peaks appear to be differentially bound between states A and B, even though only Peak 3 has differential DNA occupancy between the states (see Figure~\ref{fig:oneOverExpressedPeak}(e)). The relationship between the normalized reads for Peak 3 also does not align with the relationship between the DNA occupancy for Peak 3 when Library Size (Reads in Peaks) normalization is used. Hence, the toy example underscores that understanding the technical conditions of the selected between-sample normalization method and whether they are violated in a particular instance is crucial to ensuring accurate differential binding analysis results. 

In the Appendix, we detail frequently assumed technical conditions and connect them to common ChIP-seq between-sample normalization methods (see Table~\ref{tab:simulated-conditions} for a summary), many of which are available through the DiffBind package, which is specifically designed for identifying differentially bound peaks between sample groups \citep{DiffBind-vignette}.

\section{Simulations}
\label{sec:main-sim-results}

\begin{table}[t]
\caption{Summary of the eight unique simulation conditions. \label{tab:simulated-conditions}}
\tabcolsep=0pt
\begin{tabular*}{\textwidth}{@{\extracolsep{\fill}}>{\centering\arraybackslash}p{1cm}>
{\centering\arraybackslash}p{1cm}>
{\centering\arraybackslash}p{1cm}>
{\centering\arraybackslash}p{1.25cm}>
{\centering\arraybackslash}p{1.25cm}>
{\centering\arraybackslash}p{0.25cm}>{\centering\arraybackslash}p{0.25cm}>{\centering\arraybackslash}p{1.75cm}@{\extracolsep{\fill}}}
\makecell{Balanced} & \makecell{Equal \\ Occupancy} & \makecell{Equal \\ Background} & \makecell{\% Peaks with \\ More Occ. (A)} & \makecell{\% Peaks with \\ More Occ. (B)} & \makecell{FC \\ (A)} & \makecell{FC \\ (B)} & \makecell{Avg. \% More \\ Background/Rep$^{1}$} \\
\midrule
\checkmark  & \checkmark & \checkmark & $50$ & $50$ & $2$ & $2$ & $0$ \\
\checkmark & X & \checkmark & $50$ & $50$ & $4$ & $2$ & $0$ \\
X & \checkmark & \checkmark & $12.5$ & $87.5$ & $2$ & $2$ & $0$ \\
X  & X   & \checkmark & $12.5$ & $12.5$ & $8$ & $2$ & $0$ \\
\checkmark  & \checkmark & X & $50$ & $50$ & $2$ & $2$ & $20$ \\
\checkmark & X & X  & $50$ & $50$ & $4$ & $2$ & $20$ \\
X & \checkmark & X & $12.5$ & $87.5$ & $2$ & $2$ & $20$ \\
X  & X   & X & $12.5$ & $12.5$ & $8$ & $2$ & $20$ \\
\end{tabular*}
\begin{tablenotes}%
\item[$^{1}$] The experimental state with more background binding per replicate is chosen at random in each simulation iteration. FC = fold change in DNA occupancy when comparing non-differentially occupied peaks to peaks with more DNA occupancy in the given state. Occ. = DNA Occupancy.
\end{tablenotes}
\end{table}

To demonstrate the importance of satisfying the technical conditions of the selected between-sample normalization method on downstream differential binding analysis, we simulate ChIP-seq read count data across two experimental states under eight distinct conditions (see Table \ref{tab:simulated-conditions}). Our ChIP-seq read count data simulation primarily builds on code from Lun and Smyth \cite{csaw} as well as from Evans et al.\ \citep{Ciaran}. The code for the ChIP-seq simulations, result plots, experimental data analysis, and all toy examples is available in the GitHub repository associated with this paper.\footnote{\url{https://github.com/scolando/ChIP-Seq-norm}}

\subsection{Simulation Details}

We simulate a total of eight distinct conditions, which we summarize in Table~\ref{tab:simulated-conditions}. The eight simulation conditions cover all possible combinations of whether the three primary technical conditions -- balanced differential occupancy, equal DNA occupancy across experimental states, and equal background binding across the experimental states -- are met or violated.

Our simulations focus on the occupancy of a simple transcription factor on an artificial chromosome that is roughly 150 million base pairs long. Note that simple transcription factors often generate narrower and taller peaks than other proteins that might be of interest in ChIP-seq experiments, such as histone marks \cite{Landt2012}. Each simulation consists of two experimental states, denoted by A and B, and three replicates per experimental state (see the Appendix for simulation results with two replicates per experimental state).  

Leveraging the simple transcription factor simulation code from Lun and Smyth,\footnote{\url{https://bioinf.wehi.edu.au/csaw/}} we initialize SAM files for each replicate. Within each SAM file, we simulate 10,000 narrow peaks (with peak width of 200 base pairs) whose summit locations are approximately 15,000 base pairs apart from each other, on average, with the total proportion of peaks with differential DNA occupancy between states A and B ranging from 0.05 to 0.95 (with a step-size of 0.05). These reads are sampled from a negative binomial distribution, parametrized by the mean and dispersion \cite{csaw,Ciaran}. The mean of the negative binomial distribution represents the average amount of DNA occupancy within a given genomic region. Meanwhile, the dispersion represents the variability in the amount of DNA occupancy within a given genomic region. Like Lun and Smyth \cite{csaw}, we generate this dispersion parameter by sampling from an inverse chi-square distribution with 10 degrees of freedom. However, we vary the mean of the negative binomial distribution according to the specific simulation condition (see the simulation code for details). Once the 10,000 peaks have been simulated for each replicate, we add background reads to the SAM files by sampling from a uniform distribution (the minimum and maximum of the uniform distribution also vary according to the simulation condition). We create an input control for each replicate by sampling background reads from a uniform distribution and adding them to a new SAM file that contains no simulated peaks to model the nonspecific background binding in the replicates.

After the SAM files are created for both the replicates and their corresponding input controls, we convert them to BAM files using the xscss package \cite{csaw}\footnote{\url{xhttps://bioinf.wehi.edu.au/csaw/}}. We then perform peak calling with MACS2 on the BAM files using the parameters \texttt{-g 300734518 --mfold 2 50 --keep-dup=all -f BAM}, and supply the corresponding input control for each replicate to increase the peak calling specificity \cite{MACS}. Using DiffBind, we identify the consensus peaksets within experimental states A and B for the given simulation iteration using a minimum overlap of 75\%. We combine these peaksets such that any peak which appears in \textit{at least one} consensus peakset within an experimental state is included in the consensus peakset across states, which we use for downstream differential binding analysis. We use MACS2 as the peak caller in our simulation for two reasons. First, our focus is on normalization methods that occur after peak calling; thus, adding in different peak callers as an additional variable would make it harder to directly compare how different normalization methods affect the results. Second, the experimental datasets presented in our analysis both used MACS2 as the peak caller. Using the same peak caller for both the simulated and experimental datasets makes it easier to compare how the normalization method affects the results for both the experimental and simulated data. That said, we recognize that there are other peak callers (e.g., see \cite{suryatenggara2022}) that are suitable for analyzing broad, flat peaks, or other more complex DNA occupancy events rather than the DNA-sequence specific, high, sharp peaks for which MACS2 is optimized \cite{MACS}.

Next, we perform between-sample normalization using the DiffBind-supported between-sample normalization methods as well as MAnorm2, a package developed by Tu et al.\ \cite{MAnorm2}, which uses the \textit{common peakset} across experimental states for normalization \cite{DiffBind-vignette, MAnorm2}. For the DiffBind-supported normalization methods, we conduct our differential binding analysis in DiffBind \cite{DiffBind-vignette}. Meanwhile, we use the differential binding analysis developed in conjunction with the MAnorm2 normalization method to perform differential binding analysis on the MAnorm2-normalized read counts. For both differential binding analysis procedures, we set the false discovery rate threshold to 0.05, meaning that we only classify peaks with Benjamini-Hochberg adjusted p-values below 0.05 as differentially bound. For each simulation iteration, we return the number of true positive, true negative, false positive, and false negative differentially bound peaks associated with each normalization method, along with metadata about the simulation iteration. We iterate through this entire data-generation and differential binding analysis workflow 100 times for each combination of simulation condition and proportion of peaks with differential DNA occupancy.

To evaluate the performance of each between-sample normalization method, we devise an omniscient Oracle normalization method, which derives its size factors directly from the simulation parameters (described in more detail in the Appendix). We use the Oracle normalization method as the basis of comparison since peaks could be incorrectly identified as differentially occupied, or peaks with differential occupancy could remain undiscovered, even with perfect normalization. For example, a peak with differential DNA occupancy might not be called as a peak during peak calling due to low signal. Alternatively, a peak with true differential occupancy might have different normalized read counts between experimental states but have an adjusted p-value above the pre-specified false discovery rate (FDR) threshold due to a low number of replicates in the experimental states. 

We directly compare the sample-wide size factors generated by the Oracle to those generated by other between-sample normalization methods by computing the \textit{average absolute size factor ratio relative to the Oracle}. Let $s_{jz}$ denote the sample-wide size factor corresponding to sample $j$ for normalization method $z$, and $s_{jo}$ denote the sample-wide size factor corresponding to sample $j$ for the Oracle. Then, the absolute size factor ratio, which we denote as $\textrm{(ratio)}_{jz}$, is defined by the following piecewise function:

\begin{equation}
    \textrm{(ratio)}_{jz} = \begin{cases}
    \frac{s_{jz}}{s_{jo}} & \textrm{if }\frac{s_{jz}}{s_{jo}} \geq 1 \\
    \frac{s_{jo}}{s_{jz}} & \textrm{otherwise}
\end{cases}
\end{equation}

The average absolute size factor ratio relative, to the Oracle, for normalization method $z$ ($\textrm{(aasfro)}_{z}$) for a given simulation condition is then defined as the mean of the absolute size factor ratios for normalization method $z$ relative to the Oracle across all $n$ samples:

\begin{equation}
    \textrm{(aasfro)}_{z} = \frac{1}{n} \cdot \sum_{j = 1}^{n}\textrm{(ratio)}_{jz}
\end{equation}

We use the metric, along with the average empirical FDR and average power, to assess the relative performance of the different between-sample normalization methods in the subsequent section. 

The three simulation conditions that we assess are balanced differential occupancy, equal total DNA occupancy, and equal background bin occupancy. Balanced differential occupancy refers to the setting where the number of peaks with more DNA occupancy in state A is the same as the number of peaks with more DNA occupancy in state B (see Figure~\ref{fig:balance}). Equal total DNA occupancy refers to the setting where the total DNA occupancy in states A and B are the same (see Figure~\ref{fig:dna_bind}). Equal background bin occupancy refers to the setting where there is an equal number and distribution of rogue reads (i.e., binding in regions that are not truly occupied by the protein of interest) across states A and B (see Figure~\ref{fig:bin_obs_unit}).

\subsection{Simulation Results}
\label{sec:gen-sim-results}

\begin{table}[b]
\centering
\begin{tabular*}{\columnwidth}{@{\extracolsep{\fill}}p{3.2cm}p{5.8cm}}
Technical Condition & Normalization Methods \\
\midrule
\multirow{3}{=}{Balanced Differential DNA Occupancy} 
& TMM (Reads in Peaks) \\
& RLE (Reads in Peaks) \\
& MAnorm2 (Common Peaks) \\
\addlinespace
\multirow{2}{=}{Equal Total DNA Occupancy} 
& Library Size (Reads in Peaks) \\
& Loess Adjusted Fit (Reads in Peaks) \\
\addlinespace
\multirow{3}{=}{Equal Background Binding} 
& TMM (Background Bins) \\
& RLE (Background Bins) \\
& Library Size (Background Bins) \\
\bottomrule
\end{tabular*}
\caption{Summary of technical conditions underlying different between-sample normalization methods.}
\label{tab:technical-condition-summary}
\end{table}

\begin{figure*}[!t]%
\centering
\includegraphics[width = \linewidth, alt = {}]{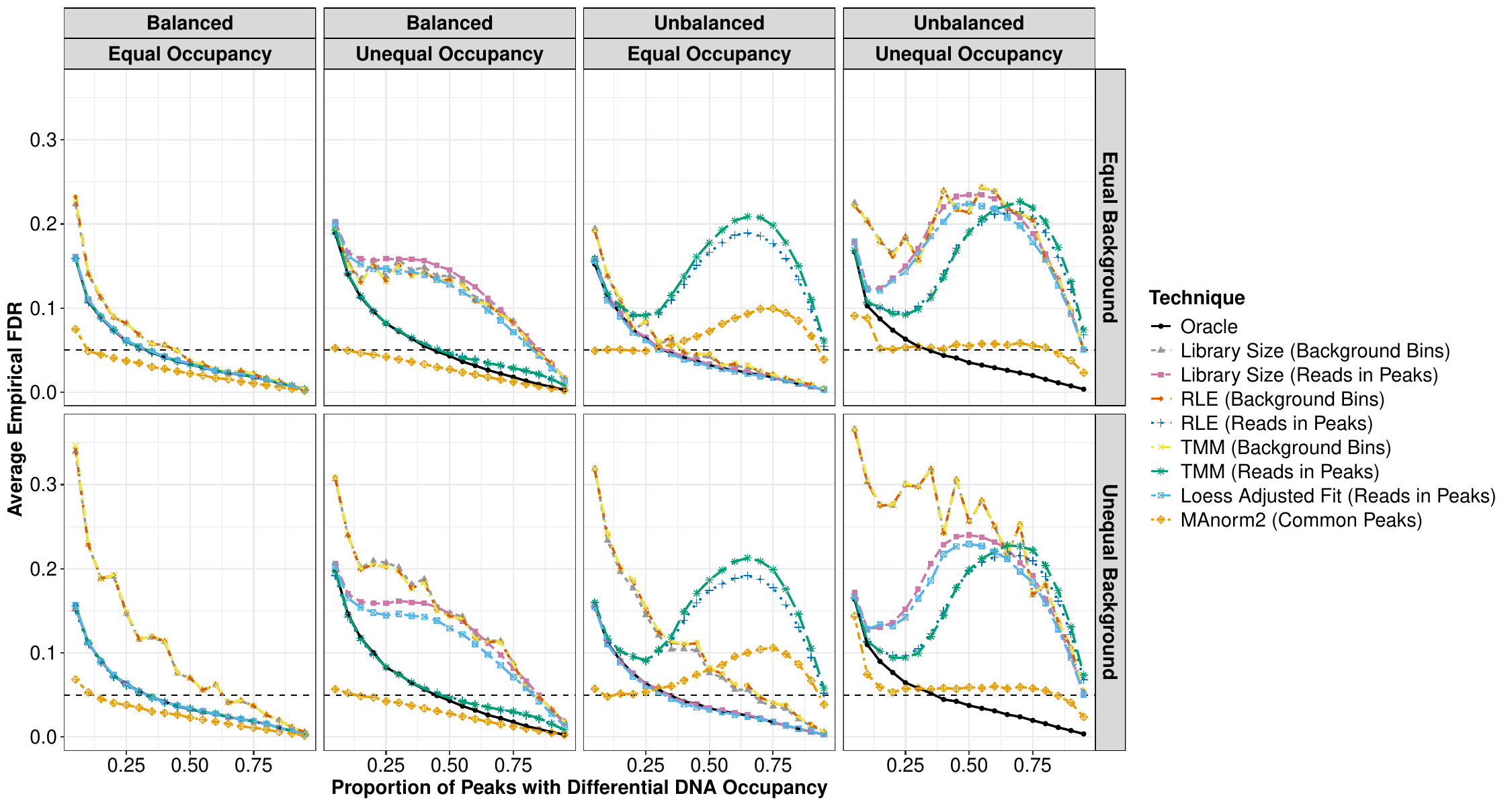}
\caption{The average empirical false discovery rate (FDR) for each normalization method, faceted by simulation condition (three replicates per experimental state and 100 simulation iterations per combination of simulation condition and proportion of peaks with differential DNA occupancy). The horizontal axis is the proportion of peaks with differential DNA occupancy, and the vertical axis is the average false discovery rate. When a curve is below the dashed horizontal black line, the associated normalization method (on average) successfully controls the false discovery rate at the pre-specified threshold of $0.05.$ We consider normalization methods that track with the Oracle (the solid black line) to be performing well with respect to the average empirical FDR in our simulation}.
\label{fig:FDR-all}
\end{figure*}

Table~\ref{tab:technical-condition-summary} connects each normalization method covered in our simulation study to the appropriate technical condition for that method.  Figure~\ref{fig:FDR-all} shows the average empirical false discovery rate (FDR) across our eight simulation conditions for each between-sample normalization method.\footnote{n.b., the average FDR will naturally decreases as the proportion of peaks with differential DNA occupancy increases since there are fewer peaks to falsely discover as differentially bound when the proportion of peaks differential occupancy between states increases.} Notably, we see that when the technical conditions of a normalization method are violated, the average empirical FDR \textit{increases}, becoming larger than the average empirical FDR associated with the Oracle -- our omniscient normalization method. In other words, a higher proportion of the peaks that are identified as differentially bound do not have true differential DNA occupancy between the experimental states when the technical conditions for the selected normalization method are violated. For example, the last two columns of Figure~\ref{fig:FDR-all} show that when the technical assumption of balanced differential DNA occupancy is violated, TMM (Reads in Peaks), RLE (Reads in Peaks), and MAnorm2 (Common Peaks) perform poorly, with their average empirical FDR diverging from that of the Oracle. Meanwhile, when there is balanced differential occupancy between the experimental states (see the first two columns in Figure~\ref{fig:FDR-all}), these normalization methods have average empirical FDR values that track closely to that of the Oracle. Likewise, between-sample normalization methods that rely on the technical condition of equal DNA occupancy between experimental states, such as Library Size (Background Bins), Library Size (Reads in Peaks) and Loess Adjusted Fit (Reads in Peaks), have an increased average empirical FDR when there is unequal DNA occupancy between the experimental states, diverging from the Oracle's average empirical FDR curve (see columns 2 and 4 of Figure~\ref{fig:FDR-all}). On the other hand, when there is equal DNA occupancy between experimental states, these methods track closely to the Oracle's average empirical FDR curve (see columns 1 and 3 of Figure~\ref{fig:FDR-all}). Finally, Figure~\ref{fig:FDR-all} demonstrates that normalization methods that use background bins rather than peaks to normalize the read counts between experimental states are affected by both unequal background binding and unequal DNA occupancy across experimental states. We hypothesize that unequal DNA occupancy impacts the average empirical FDR of between-sample normalization methods that use background bins because highly unequal DNA occupancy, particularly when there is a high proportion of peaks with differential occupancy, would cause non-trivial differences in the background binding between the experimental states.

\begin{figure*}[!ht]%
\centering
\includegraphics[width = \linewidth, alt = {}]{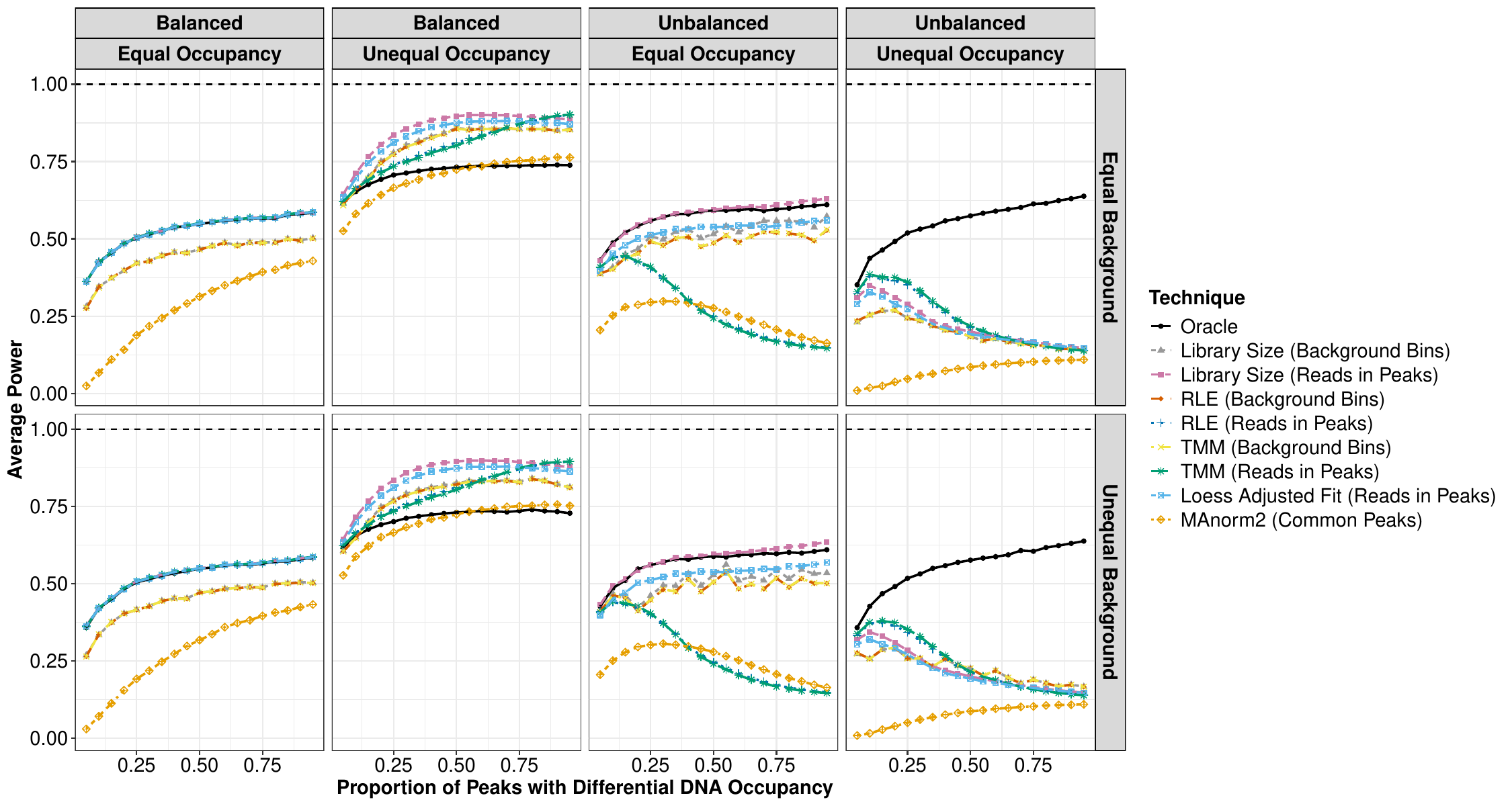}
\caption{The average power for each normalization method, faceted by simulation condition (three replicates per experimental state and 100 simulation iterations per combination of simulation condition and proportion of peaks with differential DNA occupancy). The horizontal axis is the proportion of peaks with differential DNA occupancy, and the vertical axis is the average power. The nearer a curve is to the dashed horizontal black line at $1$, the higher power the associated normalization method has (on average) for the given proportion of peaks with differential DNA occupancy.}
\label{fig:Power-all}
\end{figure*}

To further understand how violating the technical conditions underlying the selected between-sample normalization method impacts differential binding analysis results, we next examine the average power. Figure~\ref{fig:Power-all} shows that the average power associated with a normalization method's performance also tends to decrease when its technical conditions are violated. That is, peaks with differential DNA occupancy are generally less likely to be identified as differentially bound when the technical conditions for the selected normalization method are violated. For example, the average power decreases as the proportion of peaks with differential DNA occupancy increases for methods that use the mean or median peak to normalize between samples -- namely, TMM (Reads in Peaks), RLE (Reads in Peaks), and MAnorm2 (Common Peaks) -- when there is unbalanced differential occupancy between the experimental states (see the last two columns of Figure~\ref{fig:Power-all}). Thus, we would expect to have low sensitivity in downstream differential binding analysis if we were to use these normalization methods when there is unbalanced differential DNA occupancy between experimental states.

Importantly, though, Figure~\ref{fig:Power-all} highlights that the average power does not universally decrease when the technical conditions for the selected normalization method are violated. In particular, normalization methods that use background bins for normalization or that normalize using library size maintain high average power even when there is unequal total DNA occupancy between experimental states. This high power, coupled with the high average empirical FDR that we observe in Figure~\ref{fig:FDR-all}, suggests that methods that utilize the background bins or library size to normalize between states lead to low specificity in identifying peaks that are differentially bound in cases with unequal DNA occupancy between experimental states. 

In the subsequent subsections, we delve deeper into how violating each of the three technical conditions -- balanced differential DNA occupancy, equal total DNA occupancy, and unequal background binding -- affects the performance of between-sample normalization methods, as measured by the average empirical FDR, power, and absolute size factor ratio relative to the Oracle. In particular, we demonstrate how violating the technical conditions for selected normalization methods leads to incorrect between-sample normalization, which ultimately results in the misidentification of peaks with differential DNA occupancy between experimental states.

\begin{figure}[H]%
\centering
\includegraphics[width = \linewidth, alt = {Three line plots of the average empirical FDR, average power, and average absolute size factor ratio relative to the oracle associated with the between-sample normalization conditions tested when all three technical conditions are met in our simulation. The horizontal axis is the proportion of peaks with differential DNA occupancy in the simulation, ranging from 0.05 to 0.95, and the vertical axis is the value for the respective metric. In each line plot, the color of the solids denote the associated normalization method and there is a dashed black line at the notable value for the metric (0.05 for the average empirical FDR since we set that as our simulation threshold, 1.0 for the average power, and 1.0 for the average absolute size factor ratio relative to the Oracle.}]{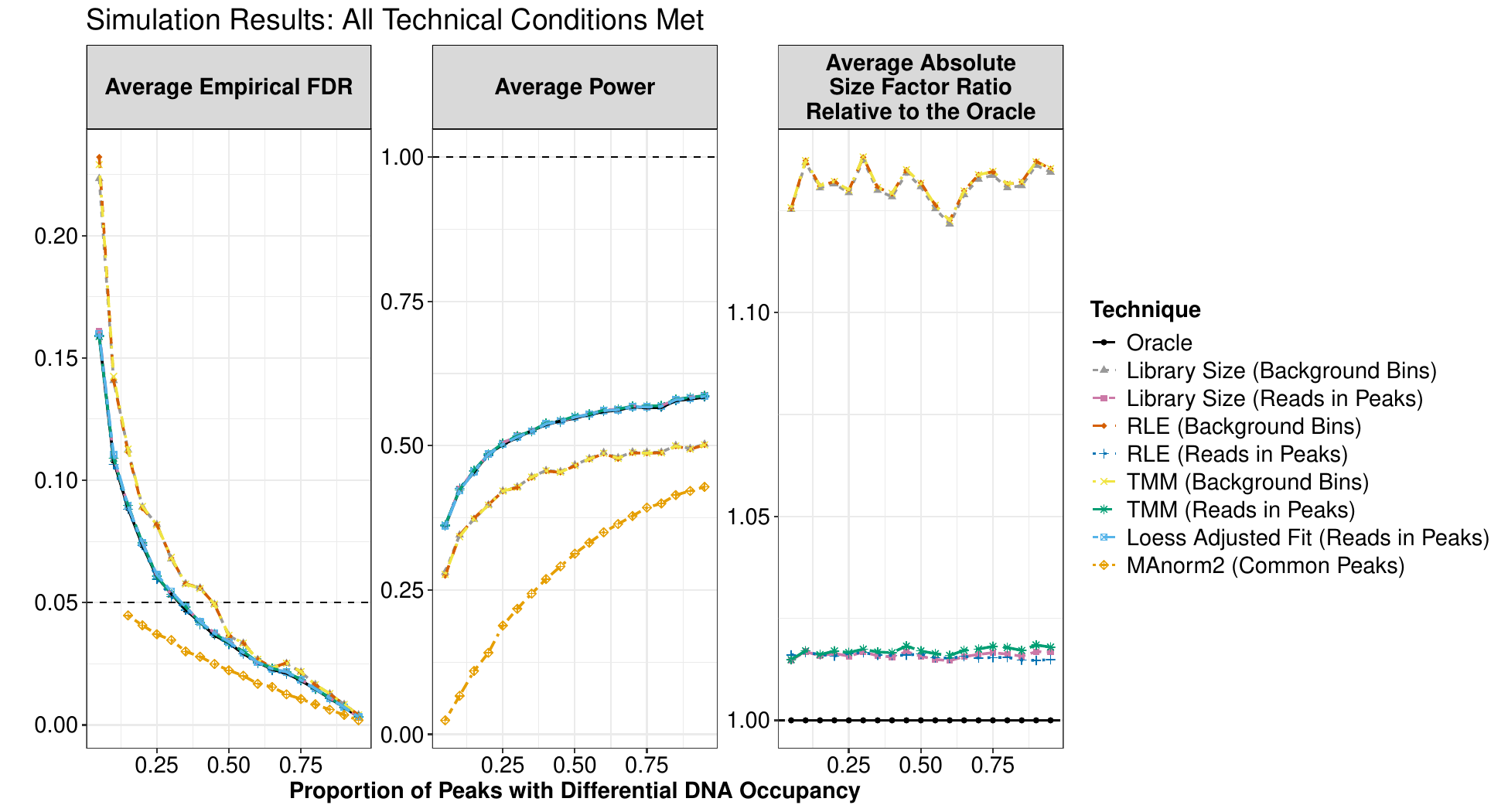}
\caption{Simulation results when all three technical conditions are met (three replicates per experimental state and 100 iterations per simulation condition and proportion of peaks with differential DNA occupancy). The horizontal axis is the proportion of peaks with differential DNA occupancy, and the vertical axis is the value for each simulation metric. The figure is faceted such that each panel represents one of the three evaluation metrics (note the scale change between each panel in the figure). The left panel displays the average empirical FDR, where the horizontal black line represents the pre-specified threshold of $0.05$. The middle panel is the average power, where the horizontal black line at $1$ represents the highest possible value. The right panel is the average absolute size factor ratio relative to the Oracle, where the horizontal line at $1$ represents the Oracle's average absolute size factor ratio with itself. A normalization method tracks closely with the Oracle if its size factor curve is close to the horizontal black line at $1$. Since MAnorm2 (Common Peaks) and Loess Adjusted Fit (Reads in Peaks) do not generate sample-wide size factors, they do not have curves in the right panel.}
\label{TTT-3-rep}
\end{figure}

\subsubsection{All Technical Conditions Met}
\label{4-rep-TTT-panel}

Per Figure~\ref{TTT-3-rep}, all normalization methods achieve an average empirical FDR close to that of the Oracle when all three technical conditions are met. Notably, MAnorm2 (Common Peaks) has an average empirical FDR lower than the Oracle's (and below the pre-specified FDR threshold of 0.05) regardless of the proportion of peaks with differential DNA occupancy. However, the middle panel of Figure~\ref{TTT-3-rep} indicates that MAnorm2 (Common Peaks) also has the uniformly lowest average power of all normalization methods, providing evidence that MAnorm2 (Common Peaks) leads to more conservative differential binding analysis than the other normalization methods we tested.

Additionally, the right panel of Figure~\ref{TTT-3-rep} shows that the average size factors corresponding to the background bin normalization methods -- RLE (Background Bins), TMM (Background Bins), and Library Size (Background Bins) -- are different from the Oracle's size factors even when all three technical conditions are met, with an (average absolute size factor ratio relative to the Oracle ranging between roughly 1.12 and 1.15). Despite the size factor discrepancy, normalization methods that use background bins still maintain an average empirical FDR and power that track closely to the Oracle's.  As such, we posit that the size factor discrepancy between normalization methods and the Oracle is due to these methods using background bins instead of peaks as the unit of normalization. 

\begin{figure}[H]%
\centering
\includegraphics[width = \linewidth, alt = {Three line plots of the average empirical FDR, average power, and average absolute size factor ratio relative to the oracle associated with the between-sample normalization conditions tested when there is unbalanced differential DNA occupancy between the two experimental states but the other two technical conditions hold in our simulation. The horizontal axis is the proportion of peaks with differential DNA occupancy in the simulation, ranging from 0.05 to 0.95, and the vertical axis is the value for the respective metric. In each line plot, the color of the solids denote the associated normalization method and there is a dashed black line at the notable value for the metric (0.05 for the average empirical FDR since we set that as our simulation threshold, 1.0 for the average power, and 1.0 for the average absolute size factor ratio relative to the Oracle.}]{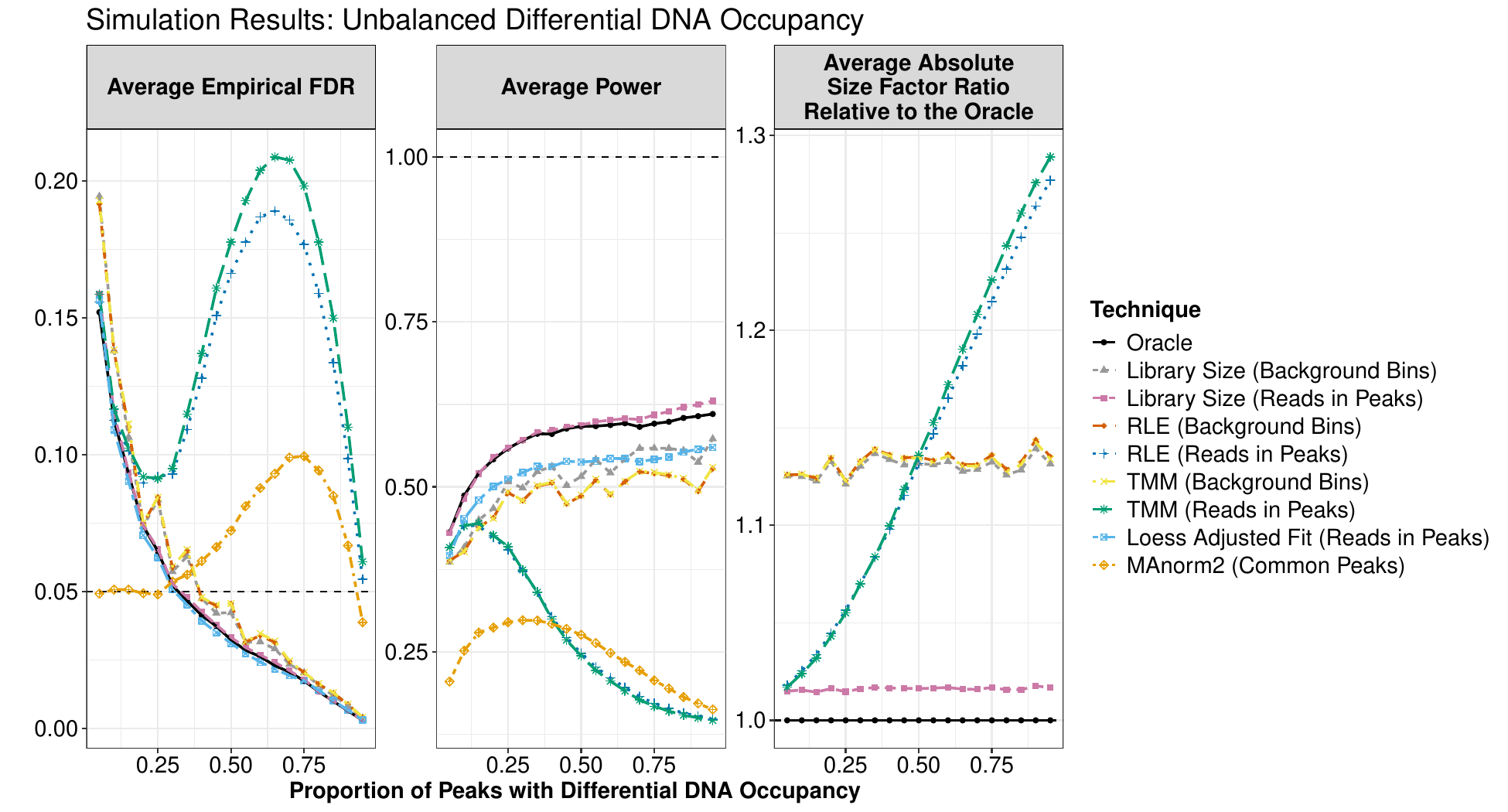}
\caption{Simulation results when only balanced differential is violated, with three replicates per experimental state. The horizontal axis is the proportion of peaks with differential DNA occupancy, and the vertical axis is the value for each simulation metric. The figure is faceted such that each subplot represents one of the three simulation metrics (note the scale change between each panel in the figure). The left panel is the average empirical false discovery rate (FDR), where the horizontal black line represents the $0.05$ FDR threshold we set. The middle panel is the average power, where the horizontal black line at $1$ represents the highest possible average power. The right panel is the average absolute size factor ratio relative to the Oracle, where the horizontal line at $1$ represents the Oracle's average absolute size factor ratio with itself. Thus, a normalization method tracks closely with the Oracle if its size factor curve is close to the horizontal black line at $1$ in the right panel. Recall that MAnorm2 (Common Peaks) and Loess Adjusted Fit (Reads in Peaks) do not generate sample-wide size factors, and so do not have curves in the right panel of the figure.}
\label{FTT-4-rep}
\end{figure}

\subsubsection{Effect of Unbalanced Differential DNA Occupancy}
\label{4-rep-FTT-panel}

When the technical condition of balanced differential DNA occupancy between experimental states is violated, but the other two technical conditions are still met, TMM (Reads in Peaks), RLE (Reads in Peaks), and MAnorm2 (Common Peaks) diverge from the Oracle across all three evaluation metrics. Indeed, as shown in the left and middle panels of Figure~\ref{FTT-4-rep}, RLE (Reads in Peaks), TMM (Reads in Peaks), and MAnorm2 (Common Peaks) all fail to control the average empirical FDR and exhibit a lower average power than the Oracle, particularly as the proportion of peaks with differential DNA occupancy increases. An explanation for this is that the peak with the mean or median read count, which these methods use to normalize between samples (see Appendix), is more likely to be a peak with differential DNA occupancy when the proportion of peaks with differential DNA occupancy increases. 

The average size factors associated with RLE (Reads in Peaks) and TMM (Reads in Peaks) also diverge from the Oracle's as the proportion of peaks with differential DNA occupancy increases (see the right panel of Figure~\ref{FTT-4-rep}). Note that MAnorm2 (Common Peaks) is not depicted in the right panel of Figure~\ref{FTT-4-rep} since it does not generate sample-wide size factors \cite{MAnorm2} The divergence in the average size factors generated by RLE (Reads in Peaks) and TMM (Reads in Peaks) compared to the Oracle verifies that these methods lead to inaccurate between-sample normalization when there is unbalanced differential DNA occupancy between states.

\begin{figure}[H]
\centering
\includegraphics[width = \linewidth, alt = {Three line plots of the average empirical FDR, average power, and average absolute size factor ratio relative to the oracle associated with the between-sample normalization conditions tested when there is unequal total DNA occupancy between the two experimental states but the other two technical conditions hold in our simulation. The horizontal axis is the proportion of peaks with differential DNA occupancy in the simulation, ranging from 0.05 to 0.95, and the vertical axis is the value for the respective metric. In each line plot, the color of the solids denote the associated normalization method and there is a dashed black line at the notable value for the metric (0.05 for the average empirical FDR since we set that as our simulation threshold, 1.0 for the average power, and 1.0 for the average absolute size factor ratio relative to the Oracle.}]{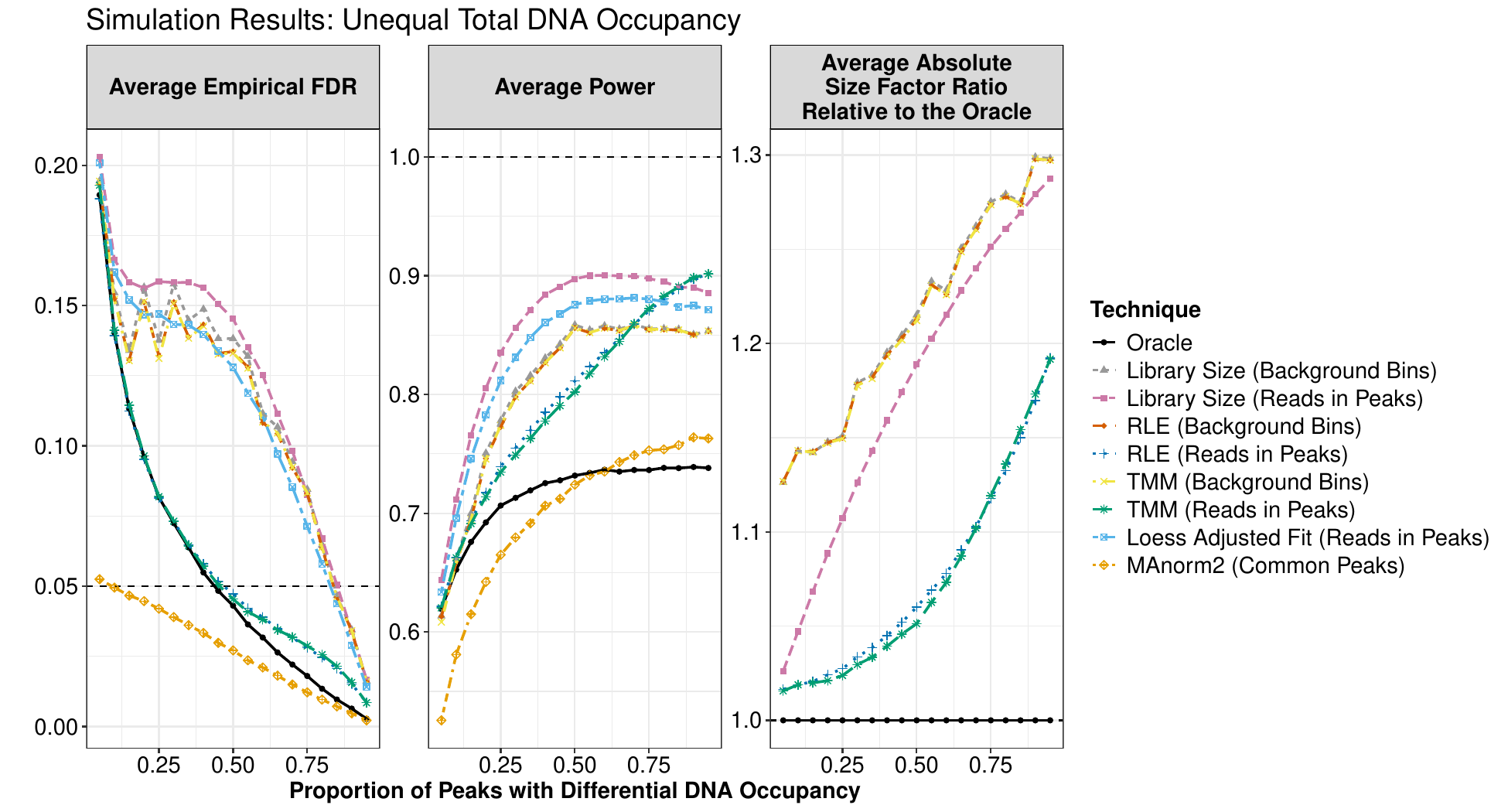}
\caption{Simulation results when only equal total DNA occupancy between experimental states is violated, with three replicates per experimental state. The horizontal axis is the proportion of peaks with differential DNA occupancy, and the vertical axis is the value for each simulation metric. The figure is faceted such that each subplot represents one of the three simulation metrics (note the scale change between each panel in the figure). The left panel is the average empirical false discovery rate (FDR), where the horizontal black line represents the $0.05$ FDR threshold we set. The middle panel is the average power, where the horizontal black line at $1$ represents the highest possible average power. The right panel is the average absolute size factor ratio relative to the Oracle, where the horizontal line at $1$ represents the Oracle's average absolute size factor ratio with itself. Thus, a normalization method tracks closely with the Oracle if its size factor curve is close to the horizontal black line at $1$ in the right panel. Recall that MAnorm2 (Common Peaks) and Loess Adjusted Fit (Reads in Peaks) do not generate sample-wide size factors and so do not have curves in the right panel of the figure.}
\label{TFT-4-rep}
\end{figure}

\subsubsection{Effect of Unequal Total DNA Occupancy}
\label{4-rep-TFT-panel}

In contrast, TMM (Background Bins), Library Size (Background Bins), RLE (Background Bins), Library Size (Reads in Peaks), and Loess Adjusted Fit (Reads in Peaks) all fail to maintain an average empirical FDR close to the Oracle's when there is unequal total DNA occupancy between experimental states (see the left panel of Figure~\ref{TFT-4-rep}). That said, in the middle panel of Figure~\ref{TFT-4-rep}, we see that all of these normalization methods still maintain a high average power. This high average power, coupled with the high average empirical FDR, indicates that TMM (Background Bins), Library Size (Background Bins), RLE (Background Bins), Library Size (Reads in Peaks), and Loess Adjusted Fit (Reads in Peaks) result in over-calling peaks as differentially bound, leading to low specificity in differential binding analysis, when their technical condition of equal DNA occupancy between states is violated.

While average absolute size factor ratio relative to the Oracle increases for all normalization methods with the proportion of peaks with differential DNA occupancy in Figure~\ref{TFT-4-rep}, normalization methods that use background bins uniformly have the largest average absolute size factor ratio relative to the Oracle. One explanation for this is that when the proportion of peaks with differential occupancy increases, so does the difference in the binding in background bins across the experimental states. As a result, the normalization methods that use background bins diverge further from the Oracle's and, as previously noted in, the average size factors for background bin normalization methods were already different from the Oracle's to begin with, likely due to using background bins rather than peaks as the unit of normalization.

\begin{figure}[H]%
\centering
\includegraphics[width = \linewidth, alt = {Three line plots of the average empirical FDR, average power, and average absolute size factor ratio relative to the oracle associated with the between-sample normalization conditions tested when there is unequal background binding between the two experimental states but the other two technical conditions hold in our simulation. The horizontal axis is the proportion of peaks with differential DNA occupancy in the simulation, ranging from 0.05 to 0.95, and the vertical axis is the value for the respective metric. In each line plot, the color of the solids denote the associated normalization method and there is a dashed black line at the notable value for the metric (0.05 for the average empirical FDR since we set that as our simulation threshold, 1.0 for the average power, and 1.0 for the average absolute size factor ratio relative to the Oracle.}]{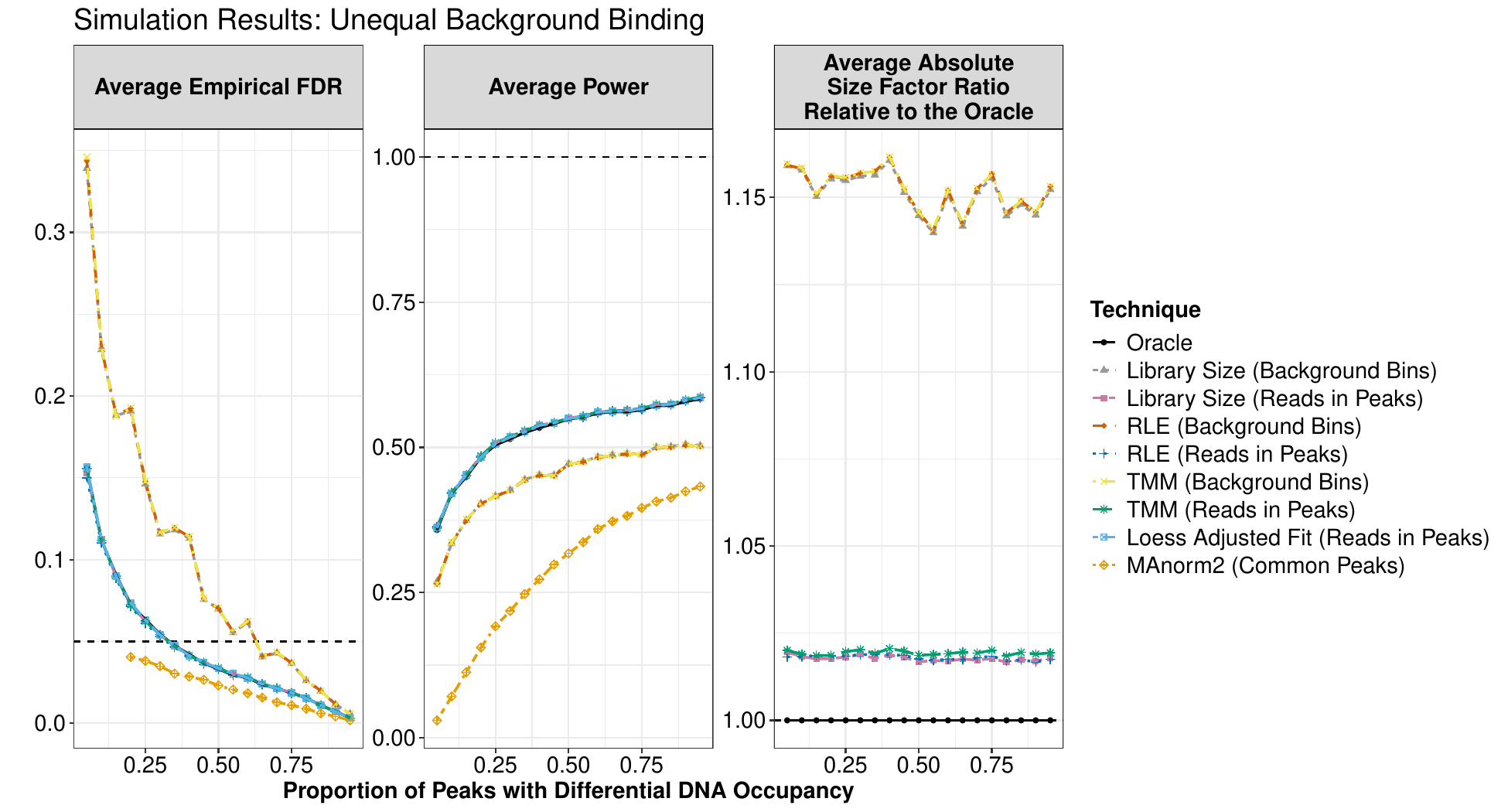}
\caption{Simulation results when only equal background binding between experimental states is violated, with three replicates per experimental state. The horizontal axis is the proportion of peaks with differential DNA occupancy, and the vertical axis is the value for each simulation metric. The figure is faceted such that each subplot represents one of the three simulation metrics (note the scale change between each panel in the figure). The left panel is the average empirical false discovery rate (FDR), where the horizontal black line represents the $0.05$ FDR threshold we set. The middle panel is the average power, where the horizontal black line at $1$ represents the highest possible average power. The right panel is the average absolute size factor ratio relative to the Oracle, where the horizontal line at $1$ represents the Oracle's average absolute size factor ratio with itself. Thus, a normalization method tracks closely with the Oracle if its size factor curve is close to the horizontal black line at $1$ in the right panel. Recall that MAnorm2 (Common Peaks) and Loess Adjusted Fit (Reads in Peaks) do not generate sample-wide size factors and so do not have curves in the right panel of the figure.}
\label{TTF-4-rep}
\end{figure}

\subsubsection{Effect of Unequal Background Binding}
\label{4-rep-TTF-panel}

Finally, when there is unequal background binding between experimental states, the between-sample normalization methods that use background bins have a higher average empirical FDR and lower average power than the Oracle (see the left and middle panel of Figure~\ref{TTF-4-rep}). Meanwhile, the average absolute size factor ratio relative to the Oracle is still approximately constant for TMM (Background Bins), RLE (Background Bins), and Library Size (Background Bins) even as the proportion of peaks with differential DNA occupancy increases (see the right panel of Figure~\ref{TTF-4-rep}). Given that the size factor ratio relative to the Oracle is roughly constant, we postulate that average empirical FDR decreases for background bin normalization methods as the proportion of peaks with differential DNA occupancy increases, not because they improve as the proportion of peaks with differential DNA occupancy increases, but instead, because there are fewer peaks to falsely discover as differentially bound.

\section{Experimental Results}
\label{sec:exp-results}

In this section, we analyze two experimental datasets (one CUT\&RUN and one ChIP-seq) to illustrate how selecting different between-sample normalization methods can affect the results of differential binding analysis, as well as the similarity between different normalization methods that rely on the same underlying technical conditions. Note that spike-in DNA is processed in the CUT\&RUN data (Example 1), we include spike-in normalization methods in our analysis. However, spike-ins are not included in the ChIP-seq data (Example 2), and therefore, we cannot incorporate spike-in normalization methods into our analysis of this dataset.

\subsection{Analysis Considerations}

While most of the normalization methods analyzed above use the exact same regions of interest, the regions of interest for MAnorm2 (Common Peaks) are computed slightly differently, and are thus not identical to those generated by the other methods. Because the simulated data consists of sharp, narrow peaks, the regions of interest for MAnorm2 (Common Peaks) and the other methods align perfectly. However, in the real data examples, the regions of interest do not align well. In particular, MAnorm2 (Common Peaks) profiles smaller regions of interest than DiffBind. This makes comparing results between MAnorm2 (Common Peaks) and other methods more difficult for experimental data because the regions of interest being compared are not the same. For this reason, we left the MAnorm2 (Common Peaks) method out of the analysis on experimental data, and focused on the  methods that use identical regions of interest to ensure a fair apples-to-apples comparison.

Another key consideration for analyzing ChIP-seq data is how to incorporate controls in the analysis pipeline. Our simulation analysis includes simulated input data that is utilized by MACS2 to determine whether a particular genomic region is called as a peak. If IgG data is available, it can also be analyzed as an experimental sample with input controls using MACS2 to identify spurious peaks. In the experimental results, both input and IgG controls are used to generate greylists that identify and filter out spurious peaks from the consensus peakset. Additionally, knockout data can be used as a control to validate peaks, since a valid peak should disappear when the gene corresponding to the protein of interest has been removed from the genome (e.g., see \cite{Krebs2014}). Thus, peaks that appear in both the wildtype and knockout samples should be eliminated as spurious. CUT\&RUN experiments do not have input data because sequences unbound by the protein of interest remain in the nucleus and are not sequenced \cite{skene2017}. In these cases, IgG data can instead be used by MACS2 (or an alternate peak calling algorithm) as the control sample, as we do here with experimental CUT\&RUN data. Once the regions of interest are identified by the peak-calling algorithm, downstream analysis focuses on determining whether the identified peaks have differential occupancy between experimental states. Input and IgG controls do not play a role in this latter part of the analysis, which determines which peaks have differential occupancy between states.

Further, researchers must determine which peak-caller algorithm to use in ChIP-seq analysis. The MACS2 algorithm, which we use for both simulated data and the experimental datasets, is optimized for identifying sequence-specific DNA-binding factors that bind to a narrow region of DNA. The MACS2 algorithm, however, also contains functionality for identifying broad peaks, which was used for the experimental CUT\&RUN data \cite{ashby2022}. For experimentalists studying histone marks that occur over broader genomic regions, a different peak-calling algorithm, such as SICER \cite{Zang2009}, that is optimized for detecting broad, low peaks, might be more suitable.

Reconciling peaks called between different replicates in an experimental state is also an important consideration in ChIP-seq analysis. In our simulations and experimental analyses, we use the narrowest region of intersection between all replicates to define the peak width for a particular peak of interest (the default in DiffBind \cite{DiffBind-vignette}). However, this could have downstream consequences for differential occupancy analysis if a large number of peaks results in a wider width for each merged peak. If visualization of peak width indicates that the width is changing significantly between experimental states, an alternate strategy such as Hidden Markov Model to identify differential occupancy might be more appropriate, for example, using the program ODIN \cite{allhoff2014}.


\subsection{Example 1: dynamic bromodomain protein 3 occupancy during African trypanosome differentiation}
\label{subsec:example1}

The first experimental dataset we analyze, from Ashby et al., characterizes the dynamic occupancy of the bromodomain protein 3 (Bdf3) in the eukaryotic protozoan parasite \textit{Trypanosoma brucei} during differentiation from bloodstream to insect forms \cite{ashby2022}. The dataset was generated by endogenously tagging a chromatin interacting bromodomain protein (Bdf3) with hemagglutinin (HA) and performing CUT\&RUN using an anti-HA antibody \citep{ashby2022, miller2023} to investigate whether Bdf3 chromatin occupancy changes as parasites transition from the mammalian bloodstream phase of the life cycle to the procyclic gut stage found in the parasite tsetse fly vector. Samples were harvested from bloodstream parasites and from several time points thereafter to produce the data. Spike-in reads from yeast DNA were added to allow for spike-in normalization during analysis. Previous analysis by Ashby et al.\ generated a high-confidence differentially bound peakset using Library Size (Spike-in), RLE (Background Bins), RPKM, and RLE (Spike-in) between-sample normalization methods \citep{ashby2022}. This analysis showed that occupancy of Bdf3 is altered at hundreds of sites as parasites transition from bloodstream to procyclic forms.  

The single-end sequencing of CUT\&RUN libraries resulted in $1.2 \cdot 10^7$ - $3.7 \cdot 10^7$ reads per library. The number of mapped reads ranged from 70-74\%, possibly due to the quality of the genome sequence. To analyze the CUT\&RUN data, raw Fastq files were trimmed for quality, and adapter sequences were removed. The resulting sequences were aligned to the Tb927v5.1 trypanosome genome using bowtie \cite{langmead2009}, requiring unique alignments using the parameters: \texttt{-best -strata -t -v 2 -a -m 1}. Spike-in reads were aligned to the yeast genome (sacCer3/R64). BAM files from the alignment were then analyzed using MACS2 in broad-peak mode to conduct peak calling on each replicate with these parameters: \texttt{-g 23650671, --keep-dup all, --nomodel, --broad}. Since there are no input controls for CUT\&RUN data (the sequences unbound by the protein of interest remain in the nucleus and are not sequenced) \cite{skene2017}, we use an IgG sample in MACS2 as the control sample to increase peak-calling specificity. The Fraction of Reads in Peaks (FRiP) scores ranged between 0.31-0.36, exceeding the ENCODE standard of $\geq$0.1 \cite{Landt2012}. The Fastq files are available in the SRA database under project number PRJNA795567.\footnote{\url{https://www.ncbi.nlm.nih.gov/sra/?term=PRJNA795567}}

To provide a simple comparison between the different between-sample normalization methods available for ChIP-seq and other similarly structured types of high-throughput data, such as CUT\&RUN, we examine the similarity in the normalized log-2 fold change between experimental states (i.e., the log-2 fold changes changes after between-sample normalization has been performed). For our analysis, we use the 3h mark as our comparison point to the bloodstream samples, as it is when the greatest change in DNA occupancy is observed relative to the bloodstream samples \cite{ashby2022}. Using MACS2 with the parameters specified above, 793 consensus peaks were identified in the bloodstream 0h samples, and 576 consensus sites were identified in the 3h differentiated samples (using a minimum overlap of three replicates). We next applied a greylist generated from the IgG control to filter out spurious peaks using the GreyListChIP package. After filtering out spurious peaks, we implemented the same between-sample normalization methods as we used in our simulation, as well as spike-in normalization methods, since spike-in DNA was also processed in the experiment. MAnorm2 was not included for the reasons cited above. Finally, we identified differentially bound regions between 0h and 3h using DiffBind (FDR $<$ 0.05).

We compare the normalized log-2 fold changes between experimental states for each between-sample normalization method using Principal Component Analysis (PCA) (Figure~\ref{experimental-results-pca}).  An UpSet plot of the differentially bound peaksets was also generated with each between-sample normalization method 
(Figure~\ref{experimental-results-UpSet}). Since we cannot display every possible distinct intersection of the ten differentially bound peaksets in the UpSet plot, Figure~\ref{experimental-results-UpSet} provides information on the twenty largest groups of distinct overlaps between the differentially bound peaksets. 

The PCA results support our simulation findings that between-sample normalization methods with the same technical conditions yield very similar normalization outcomes. Notably, the first principal component, which explains nearly all of the variability in the normalized log-2 fold changes (99.45\%), distinguishes peak-based normalization methods from those that use spike-ins or background bins for between-sample normalization. Within the peak-based normalization methods, Loess Adjusted Fit (Reads in Peaks) and Library Size (Reads in Peaks) produce extremely similar normalized log-2 fold changes (see Figure~\ref{experimental-results-pca}). This is expected since both normalization methods require an equal amount of total DNA occupancy between states, as we observed in our simulations (see Figure~\ref{TFT-4-rep}). The second principal component, which explains 0.40\% of the variability in the normalized log-2 fold changes, separates the background bin normalization methods from spike-in normalization methods. The primary technical conditions for spike-in normalization methods are that the spike-in DNA comes down in the immunoprecipitation at the same frequency regardless of the experimental states and that experimental artifacts have the same effects on spike-in DNA and the experimental samples. On the other hand, background bin normalization methods rely on equal background binding across experimental states for accurate normalization (see Figure~\ref{TTF-4-rep}).

{In total, between 270 and 457 peaks were called as differentially bound for each normalization method in this dataset, which verifies our simulation results that the choice of between-sample normalization has a substantial impact on which peaks are identified as differentially bound, especially when not every technical condition is satisfied. One hundred and seventy peaks, or between 38\% and 63\% of the differentially bound peaks found by each method, were identified as differentially bound by every selected normalization method.  Leveraging the language from Ashby et al., these 170 peaks in the distinct intersection between all normalization methods constitute a high-confidence peakset, as they were found to have differential DNA binding between the 3h differentiation and the 0h bloodstream state regardless of the between-sample normalization method used. As Ashby et al.\, emphasize, the choice of normalization method has an ``outsized effect" on which regions are identified as differentially bound between experimental states. The high-confidence peakset circumvents this problem by containing only differentially bound peaks that are robust to the choice of between-sample normalization method \citep{ashby2022}. 

\begin{figure}[H]%
\centering
\includegraphics[width = 0.9\linewidth, alt = {Scatterplot of the tested normalization methods across the first two principal components using data from Ashby et al. The principal components analysis is based on the similarity of the log-2 fold changes of the 3h timepoint sample compared to the bloodstream sample after between-sample normalization was performed on the data.}]{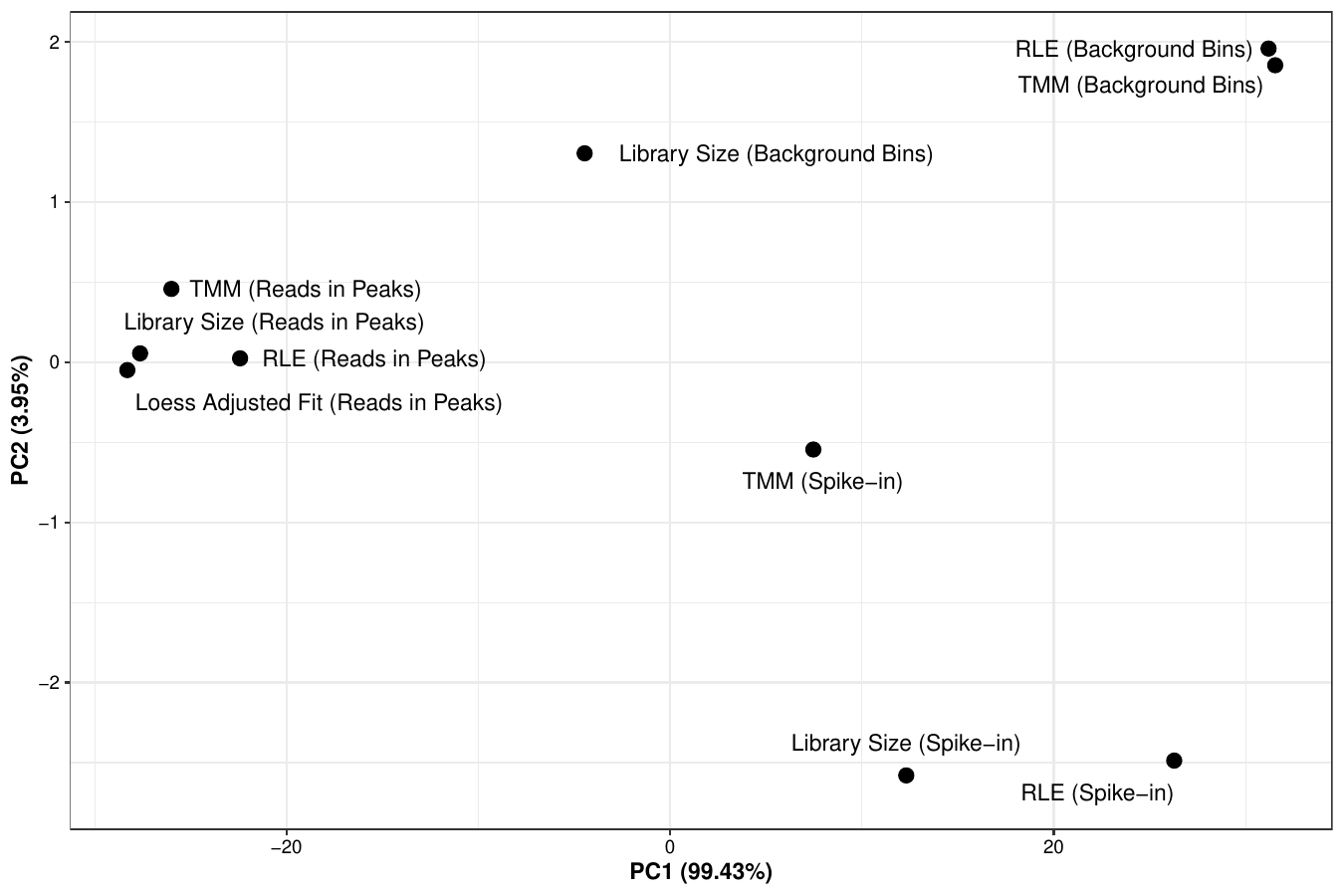}
\caption{PCA of the log-2 fold changes after between-sample normalization has been performed between the bloodstream parasites and those induced to transition into procyclic forms by the 3h incubation. The horizontal axis corresponds to the first principal component, and the vertical axis corresponds to the second principal component. Almost all the variability in the log-2 fold changes after between-sample normalization is explained by the first principal component (99.43\%), and 0.40\% of the variability is explained by the second principal component. Notably, the first principal component primarily separates methods that use peaks to normalize across states from methods that use spike-ins or background bins to normalize across states. Meanwhile, the second principal component appears to distinguish the background bin normalization methods from the spike-in normalization methods.}
\label{experimental-results-pca}
\end{figure}

\vfill
\newpage

\begin{figure}[H]%
\centering
\includegraphics[width = \linewidth, alt = {UpSet plot that compares the differential binding peaksets generated using the different tested between-sample normalization methods on the data from Ashby et al.}]{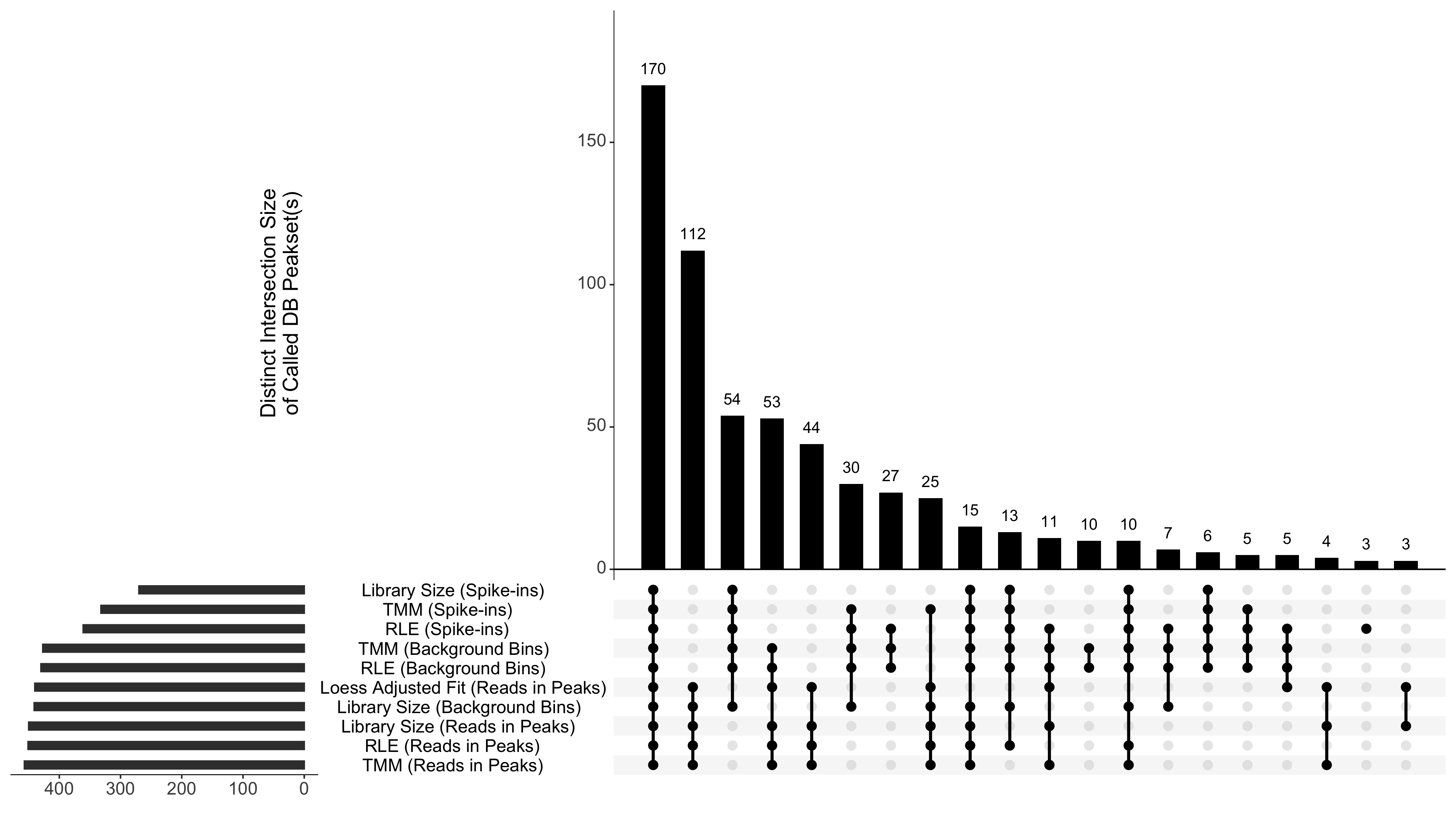}
\caption{UpSet plot comparing the peaks identified as differentially bound after selecting different between-sample normalization methods. The plot visualizes the 20 largest distinct intersections. The number of peaks identified as differentially bound using each normalization method is given in the horizontal bar plot next to the normalization method's name. One hundred and seventy peaks were identified as differentially bound by all of the ten between-sample normalization methods that we tested.}
\label{experimental-results-UpSet}
\end{figure}

\subsection{Example 2: differential oestrogen receptor-$\alpha$ binding in breast cancer cell lines}

For our second experimental analysis, we use a subset of ChIP-seq data from Ross-Innes et al., which profiles oestrogen receptor-$\alpha$ (ER) binding in tamoxifen-resistant ER-positive cell lines (in T-47D, ZR75-1, and MCF-7 tissues) versus tamoxifen-responsive ER-positive cell lines (in TAM-R and BT-474 tissues) for breast cancer patients with distinct clinical outcomes \cite{ross-innes2012}. As previously mentioned, this dataset does not include spike-in controls. In our analysis of the data, we aim to identify genomic regions with differential occupancy of the ER transcription factor between tamoxifen-resistant and tamoxifen-responsive cell types. Specifically, we want to detect differentially bound genomic regions that are robust to tissue-specific variability. So, we aggregate replicates from the different tissue types into the broader categories of tamoxifen-resistant and tamoxifen-responsive cell lines for our analysis.

Each cell line has at least two replicates and a corresponding input sample, yielding four tamoxifen-resistant replicates (2 TAM-R and 2 BT-474) and seven tamoxifen-responsive replicates (2 T-47D, 2 ZR75-1, and 3 MCF-7). Ross-Innes et al.\ processed the cell line material by cross-linking the proliferating cells using an anti-ER (SC-543) antibody and following the ChIP protocol described in Schmidt et al.\ \cite{ross-innes2012, Schmidt2009}. The resulting sequences were then processed with the Illumina analysis pipeline (version 1.6.1) and aligned to the hg19 Human Reference Genome (NCBI build 36.1, March 2008) using BWA (version 0.5.5) \cite{ross-innes2012}. The single-end sequencing of the ChIP-seq libraries resulted in roughly $2.2 \cdot 10^7 - 8.3 \cdot 10^7$ reads per library. Additional information on how the experimental data was processed is available in Ross-Innes et al.\ \cite{ross-innes2012} and in the GEO database under accession GSE32222,\footnote{\url{https://www.ncbi.nlm.nih.gov/geo/query/acc.cgi?acc=GSE32222}} the Fastq files for the full data across all chromosomes are also available under the same accession number.

To simplify our analysis, we use the subset of the sequences that were aligned to chromosome 18 in the hg19 Human Reference Genome. The chromosome 18 observations were processed for the DiffBind vignette and are accessible through the DiffBind package \cite{DiffBind-vignette}. The peaks in each replicate were identified using MACS2 on the BAM files with the parameters: \texttt{-g 2.7e9 --nomodel --extsize 137 --bw 300 --mfold 5 20 -p 1e-5 -f BAM}. The corresponding input sample for each cell line was also supplied to MACS2 in order to increase peak-calling specificity \cite{MACS}. After peak calling, the peaks were then subsetted to only those on chromosome 18. Six out of seven FRiP scores for the subsetted data exceeded the ENCODE standard of $\geq$0.1 (ranging from 0.10 to 0.31). However, one replicate from the TAM-R tissue had a FRiP score of 0.06, which is below the ENCODE standard. A blacklist based on the hg19 Human Reference Genome and greylists based on the input controls were next created via DiffBind to further cull the set of peaks considered in downstream analysis. After applying the blacklist and greylists, 1,336 consensus peaks were identified in the tamoxifen-resistant cell line samples, and 2,453 consensus peaks were identified in the tamoxifen-responsive cell line samples (at least two replicates within an experimental state had to identify the region as significantly enriched with DNA-protein binding for it to be considered a consensus peak within the experimental state). Taking the union of these peaksets yielded 2,698 consensus peaks across the experimental states. We then applied the same between-sample normalization methods as we used in our simulation (except for MAnorm2, for reasons mentioned above) to normalize the read counts in each consensus peak. Finally, we determined which genomic regions were differentially bound using DiffBind (FDR $<$ 0.05) \cite{DiffBind-vignette}.

Like we see in the CUT\&RUN experimental analysis, the between-sample normalization methods that rely on the same technical conditions generate normalized log-2 fold changes that are similar to one another. Indeed, the first principal component, which explains 91.23\% of the variability in the normalized log-2 fold changes, separates background bin normalization methods from the peak-based normalization methods (see Figure~\ref{fig:chipseq-PCA}). The peak-based normalization methods that rely on the technical condition of equal total DNA occupancy between states -- Library Size (Reads in Peaks) and Loess Adjusted Fit (Reads in Peaks) -- are distinguished from those that rely on the balanced differential DNA occupancy -- TMM (Reads in Peaks) and RLE (Reads in Peaks) by the second principal component, which accounts for 4.91\% of the variability in the normalized log-2 fold changes.

Between 245 and 440 peaks were called as differentially bound for each normalization method for this dataset, with 209 peaks being considered high-confidence since they were identified as differentially bound regardless of the selected between-sample normalization method. Between 48\% and 85\% of peaks found by each individual method were considered high-confidence differentially bound (see Figure~\ref{fig:chipseq-UPset}). This result has a broader range than the 38\%-63\% of peaks found with each individual method that were classified as high-confidence in the trypansome dataset (see Figure~\ref{experimental-results-UpSet}). However, for both datasets, high-confidence peaks capture at least the top 38\% (with most methods having greater than 50\% of their differentially bound peakset belonging to the high-confidence peakset) of all peaks identified as differentially bound by the various between-sample methods in the two experimental datasets analyzed.

\begin{figure}[H]
    \centering
    \includegraphics[width=0.9\linewidth]{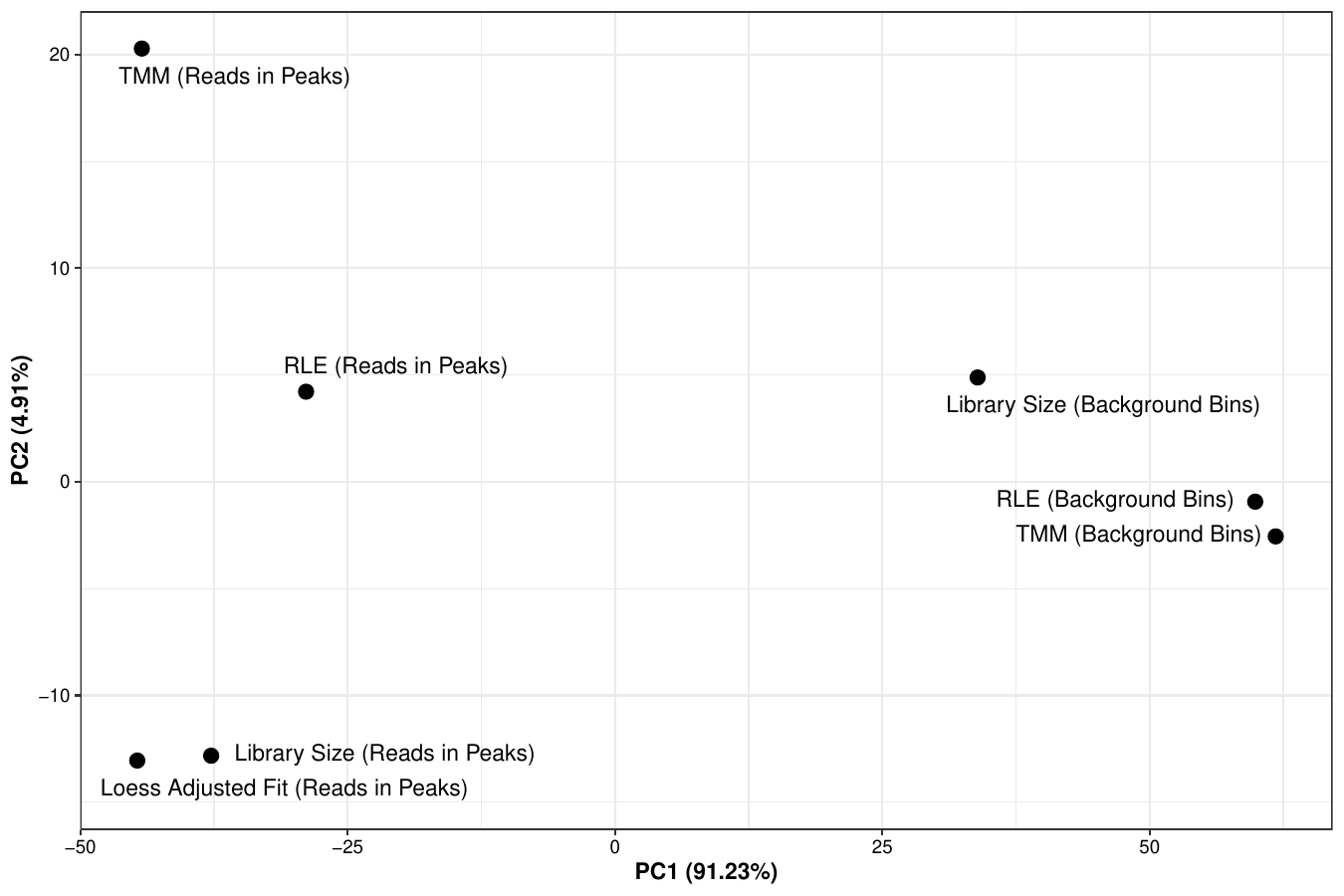}
    \caption{PCA of the log-2 fold changes after between-sample normalization has been performed between the tamoxifen-resistant and tamoxifen-responsive cell lines. The horizontal axis corresponds to the first principal component, and the vertical axis corresponds to the second principal component. A majority of the variability in the log-2 fold changes after between-sample normalization is explained by the first principal component (91.23\%), and 4.91\% of the variability is explained by the second principal component. Notably, the first principal component primarily separates methods that use peaks to normalize between states from methods that use spike-ins or background bins to normalize between states. Meanwhile, the second principal component appears to distinguish the reads in peak normalization methods that rely on the technical condition of equal DNA occupancy between experimental states (Library Size (Reads in Peaks) and Loess Adjusted Fit (Reads in Peaks) from those that rely on balanced differential DNA occupancy between states (TMM (Reads in Peaks) and RLE (Reads in Peaks)).}
    \label{fig:chipseq-PCA}
\end{figure}

\begin{figure}[H]
    \centering
    \includegraphics[width=\linewidth]{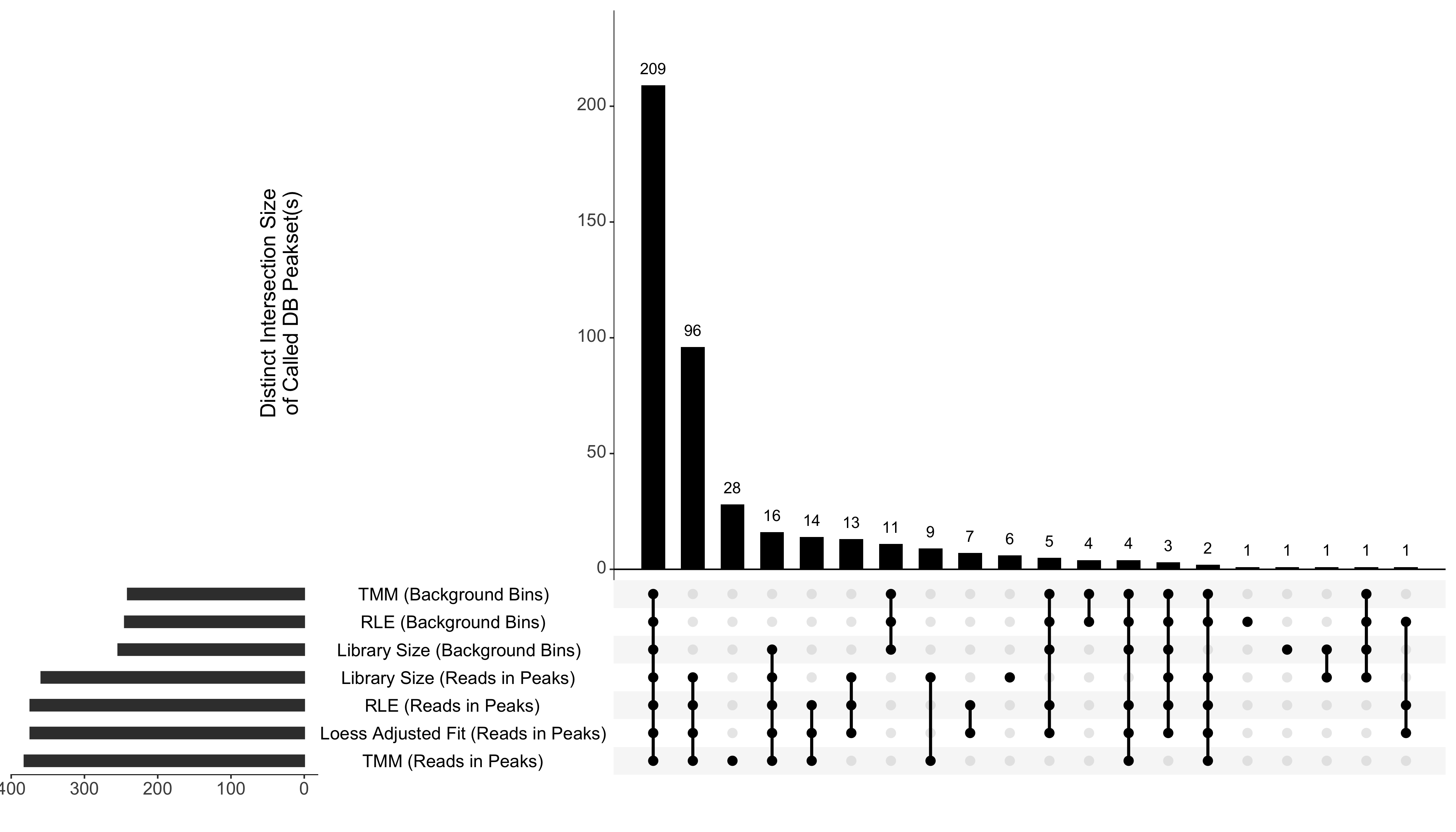}
    \caption{UpSet plot comparing the peaks identified as differentially bound between the tamoxifen-resistant and tamoxifen-responsive cell lines after selecting different between-sample normalization methods. The plot depicts the 20 largest distinct intersections. The number of peaks identified as differentially bound using each normalization method is given in the horizontal bar plot next to the normalization method's name. 209 peaks were identified as differentially bound by all of the seven between-sample normalization methods that we tested.}
    \label{fig:chipseq-UPset}
\end{figure}

\section{Discussion}

Our results show that when performing differential binding analyses, the preceding between-sample normalization step is crucial to get right as the technical conditions underlying the selected between-sample normalization method often drive the accuracy of the downstream differential binding analysis. Indeed, our simulation results demonstrate that the performance of differential binding analysis declines when the technical conditions for the selected between-sample normalization method are violated, as measured by the average empirical false discovery rate, power, and sample-wide size factors relative to the omniscient Oracle method. The experimental data externally validate our simulation results. In particular, normalization methods that rely on the same technical conditions cluster together in the PCA plot of the normalized log-2 fold changes in both experimental analyses.

In some experiments, a researcher will have a sense of whether and how the technical conditions underlying a particular between-sample normalization method are met. For example, if one experimental state causes cell damage or death, the researcher might expect an increase in chromatin shearing, and with that, a potential increase in background binding within that specific experimental state. Alternatively, if one experimental state makes it difficult to obtain sufficient numbers of cells to begin the experiment, there could also be an increase in the amount of noise, or background binding, in that particular experimental state. There is also sometimes evidence of global changes in DNA occupancy, such as with a global decrease in H3K27me3 levels \cite{agostinho2023}, or a global decrease in CTCF occupancy during the transition from pluripotency to early neuronal lineage commitment for neonatal mouse brains \cite{beagan2017}.

In cases where the researcher does not have a preconceived notion of any potential technical condition violations, a potential analysis might include implementing multiple between-sample normalization methods on the data, like Ashby et al.\, to determine a high-confidence differentially bound peakset that is less sensitive to one's choice of between-sample normalization method \cite{ashby2022}. If multiple normalization methods yield the same set of differentially bound peaks, there is strong evidence that the differential binding is the result of genuine differences in DNA occupancy between the experimental states, rather than the consequence of experimental or statistical artifacts. In the experimental data we analyzed, roughly half of the called peaks were called as differentially bound for every normalization method. In other words, the high-confidence peakset is made up of roughly half of the differentially bound peaks for each method.

Our analysis in this paper focuses on the differential binding analysis of transcription factors, which typically produce sharp, narrow peaks \cite{Landt2012}. However, other popular proteins of interest in ChIP-seq experiments, such as histone marks, are characterized by broader, lower-intensity peaks \cite{Landt2012}, or other more complex occupancy behavior. In principle, the same downstream methods described in this paper for detecting differentially bound regions after peak calling could be used following the peak-calling step for such proteins. However, because our simulations did not model data with a broad and low genomic distribution, it is possible that some of the technical condition violations explored here might not apply in exactly the same way. An analysis of between-sample normalization methods and the impact of relevant technical conditions in cases with broader, lower intensity occupancy could be an avenue for future work that builds on our study.


\section{Competing interests}
No competing interest is declared.

\section{Author contributions statement}
S.C. conducted the computational study and wrote the manuscript. D.S. ran the biological experiment, provided experimental data, and wrote \& reviewed the manuscript. J.H. conceived of the study, supported the computational work, and reviewed the manuscript.

\section{Acknowledgments}
The authors thank the anonymous reviewers for their valuable suggestions. 
S.C. was supported in part by grants from the Pomona College SURP program and Kenneth Cooke Summer Research Fellowship. D.S. was supported in part by NSF CAREER grant 2041395. J.H. was supported in part by NIH GM112625. 

\raggedright
\bibliographystyle{plain}
\bibliography{reference.bib}




\newpage

\section{Appendix}

\subsection{Normalization Methods and their Technical Conditions}
\label{sec:norm-methods}

In this section, we identify the primary technical conditions underlying ChIP-seq between-sample normalization methods and categorize common ChIP-seq between-sample normalization methods based on the technical conditions we identify. Before delving into the normalization methods and their respective technical conditions, though, we define two relevant terms:

\begin{enumerate}
    \item \textit{Reads in Peaks Normalization} methods only use reads from the consensus peakset across experimental states \citep{DiffBind-vignette}. Figure \ref{fig:peak_obs_unit} provides an illustration of the normalization unit for Reads in Peaks normalization methods.

\begin{figure}[H]
    \centering
    \includegraphics[width = \linewidth, alt = {}]{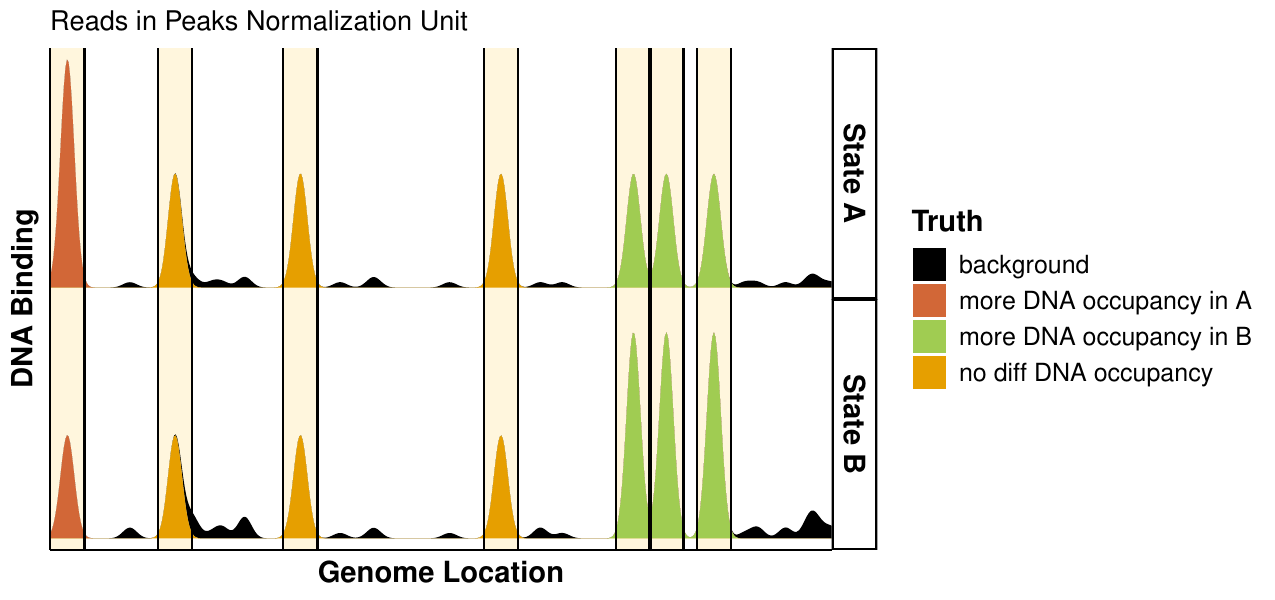}
    \caption{Reads in Peaks' normalization unit. The vertical axis represents the amount of DNA binding, and the horizontal axis represents the location of the DNA binding in the genome. The plots are faceted based on their experimental state. The genomic regions highlighted in yellow are normalization units for the Reads in Peaks methods. The black vertical lines denote the boundaries of each unit. Notice that only the peaks are used in the Reads in Peaks normalization methods.}
    \label{fig:peak_obs_unit}
\end{figure}

\begin{figure}[H]
    \centering
    \includegraphics[width = \linewidth, alt = {}]{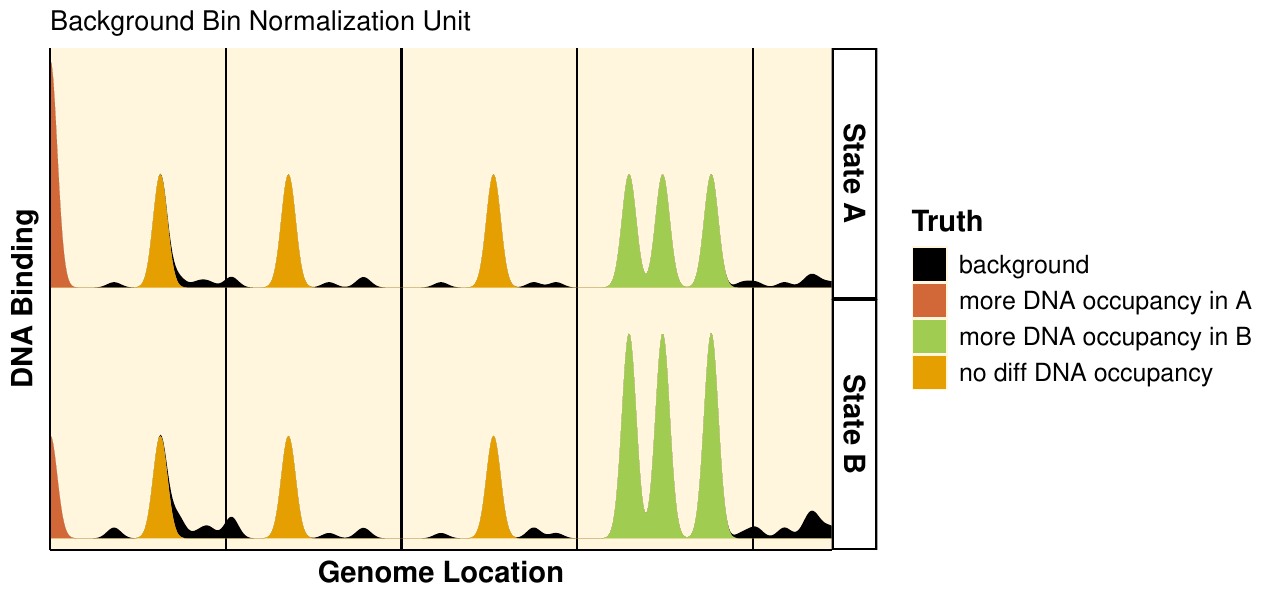}
    \caption{Background Bins' normalization unit. The vertical axis represents the amount of DNA binding, and the horizontal axis represents the genome location of the DNA binding. The plots are faceted based on their experimental state. Every genomic region highlighted in yellow would be used in Background Bins Normalization. The black vertical lines denote the boundaries of each normalization unit.}
    \label{fig:bin_obs_unit}
\end{figure}

\newpage

    \item \textit{Background Bins Normalization} methods partition each chromosome (that contains a peak in the consensus peakset) into roughly 15,000 base pairs long bins. These bins are used to normalize between experimental states \citep{DiffBind-vignette}. Figure~\ref{fig:bin_obs_unit} provides an illustration of the normalization unit for Background Bins normalization methods.
\end{enumerate}

\subsubsection{Normalization by Library Size}
\label{library-size-norm-methods}

Normalization by Library Size methods aim to remove differences in DNA binding that are due to experimental artifacts, rather than differential DNA occupancy between samples, by scaling each sample's raw reads by the total amount of DNA binding in that sample.

\paragraph{Technical Conditions}
\label{subsec:library-technical-conds}
\begin{itemize}
\item Equal Amount of Total DNA Occupancy: The total amount of DNA occupancy is equal across the experimental states. That is, each experimental state has the same amount of total DNA occupancy per cell (see Figure~\ref{fig:dna_bind}).
\end{itemize}

\begin{figure}[H]
    \centering
    \includegraphics[width=\linewidth, alt = {}]{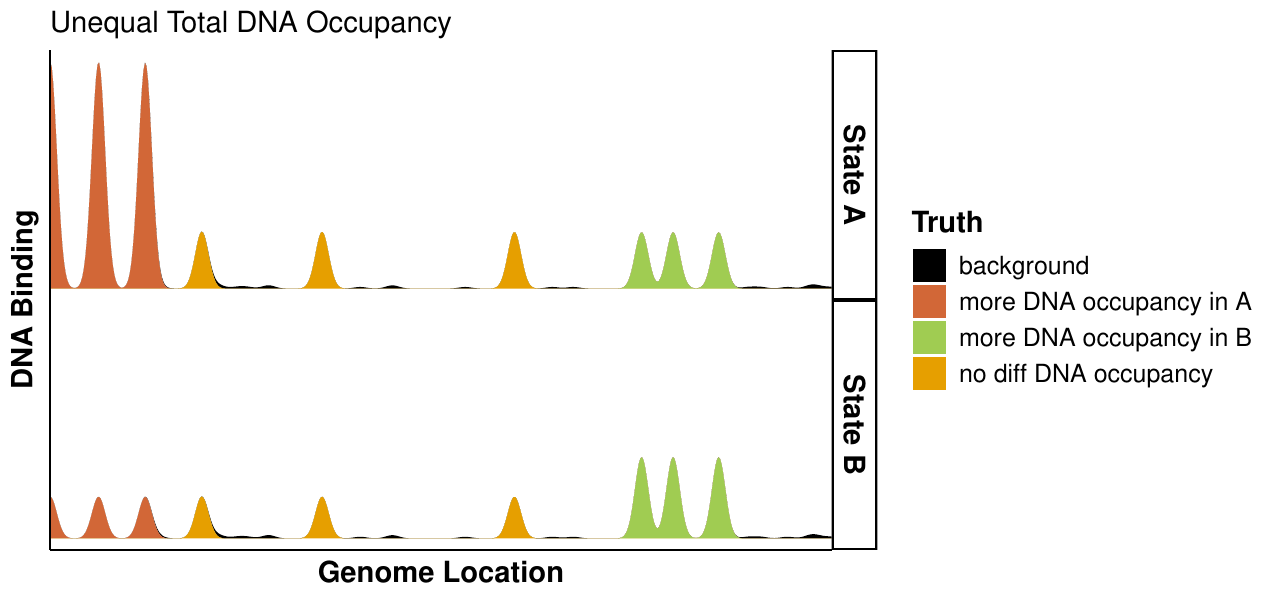}
    \caption{An illustration showing balanced differential DNA occupancy and constant background but {\bf unequal} total DNA occupancy across the experimental states, with state A having more total DNA occupancy than state B.}
    \label{fig:dna_bind}
\end{figure}

\paragraph{Methods}

\begin{enumerate}
\item Library Size (Reads in Peaks) first sums the raw read counts for peaks in the consensus peakset across experimental states to calculate the total number of raw reads in the peaks for the given sample. The sample is then normalized between samples by dividing the raw read count in each peak by the sample's total raw read count in the peaks \citep{TotalCount}. Hence, the size factor for sample $j$ is equivalent to sample $j$'s total raw peak read count under Library Size (Reads in Peaks). That is:

\begin{equation}
    s_j = \sum_{i = 1}^{I}k_{ij}
\label{eqn:lib_RIP}
\end{equation}

\noindent \textit{where $I$ is the number of consensus peaks across the experimental states and $k_{ij}$ is the raw read count associated with consensus peak $i$ in sample $j$.} \newline

\item Loess Adjusted Fit (Reads in Peaks) is a normalization method native to csaw package \cite{csaw-package} that works similarly to the variant of fast linear loess introduced by Ballman et al.\ \cite{loess} and adjusts for the sample's library size and mean centering when normalizing between sample groups \cite{DiffBind-vignette}. Loess Adjusted Fit (Reads in Peaks) works by normalizing the raw read counts via local regression on the MA values, meaning there are no sample-wide size factors calculated for Loess Adjusted Fit (Reads in Peaks). Here, the M value for a peak is the difference in the log-transformed DNA binding within the given peak across two samples, and the A value is the average log-2 binding level for that peak across the two samples \cite{MA-values}. Formally, the M and A values between two samples, $j$ and $l$, are defined as follows for some consensus peak $i$:

\begin{equation}
    M_i = \log_2(k_{ij}) - \log_2(k_{il})
\end{equation}

\begin{equation}
    A_i = \frac{1}{2}\biggl(\log_2(k_{ij}) + \log_2(k_{il})\biggl)
\end{equation}

\noindent\textit{where $k_{ij}$ is the raw read count associated with consensus peak $i$ for sample $j$, and $k_{il}$ is the raw read count associated with consensus peak $i$ for sample $l$.} \newline

Loess Adjusted Fit (Reads in Peaks) normalization employs the following general algorithm \cite{loess}:
\begin{enumerate}
    \item[(a)] Construct an average array that consists of the average log-2 read counts for each consensus peak across the experimental states.
    \item[(b)] Plot a modified MA plot between the raw read counts in sample $j$ and the average array from step (a).
    \item[(c)] Fit a loess curve on the modified MA plot.
    \item[(d)] Subtract the loess-fitted value from the relevant modified $M$ value to compute the normalized read count estimates for sample $j$. 
    \item[(e)] Repeat steps (b) through (d) for each sample across the experimental states.
    \item[(f)] Repeat steps (a) through (e) until the algorithm converges for each peak $i$ and sample $j$ across the experimental states. These converged estimates would be the normalized read counts used in downstream differential binding analysis.
\end{enumerate}  

Let $\mathrm{mean}_i(\log_2(k_i))^t$ denote the average array calculated in step (a) for the algorithm iteration $t$. Then, the $M$ and $A$ values in the modified MA plot between the sample $j$ and the average array are computed as follows for algorithm iteration $t$: 

\begin{equation}
    M_{i}^t = \log_2(k_{ij}) - \mathrm{mean}_i(\log_2(k_i))^t
\end{equation}

\begin{equation}
    A_{i}^t = \mathrm{mean}_i(\log_2(k_i))^t
\end{equation}

Next, let $\mathrm{loess}(\mathrm{mean}_i(\log_2(k_i))^t)$ denote the loess-fitted value at the consensus peak $i$ that is computed in step (c) of the normalization algorithm. For step (d), this fitted value is subtracted from $M_i^t$ to calculate the normalized read count estimate, $k_{ij}^{t*}$, for peak $i$ in sample $j$ for algorithm iteration $t$:

\begin{equation}
    k_{ij}^{t*} = M_{i}^t -\mathrm{loess}(\mathrm{mean}_i(\log_2(k_i))^t)
\end{equation}

In step (e), the process of calculating the normalized read count estimate, $k_{ij}^{t*}$, is then repeated for each peak and sample across the experimental states. In step (f), the entire process (steps (a) through (e)) is repeated until the algorithm converges to a singular normalized read count estimate for each consensus peak and sample, meaning that the normalized read count for peak $i$ in sample $j$ for Loess Adjusted Fit (Reads in Peaks) is the following:

\begin{equation}
    k_{ij}^* = \lim_{t \to \infty}\bigg( M_i^{t} - \mathrm{loess}(\mathrm{mean}_i(\log_2(k_i)^t))\bigg)
\end{equation}

As explained by Lun and Smyth \cite{csaw-package} as well as Stark and Brown \cite{DiffBind-vignette}, Loess Adjusted Fit (Reads in Peak) normalization can correct trended biases in the data, where the fold change (M values) systematically change as the average binding levels (A values) change. However, if these systematic changes are due to genuine biological signal, then normalizing the data with Loess Adjusted Fit (Reads in Peaks) could lead to peaks with differential DNA occupancy between experimental states going
undetected or conversely, for peaks without differential DNA occupancy between experimental states to be identified as differentially bound during differential binding analysis \citep{DiffBind-vignette}. Crucially, when there is unequal DNA occupancy between the experimental states, we expect a systematic difference in the fold changes as the average binding levels change because of the different total DNA occupancy between states. Hence, Loess Adjusted Fit (Reads in Peaks) relies on the technical condition that there is an equal DNA occupancy between experimental states to correctly normalize between sample groups.

\end{enumerate}

\paragraph{Motivation for Library Methods} 
When there is an equal amount of total DNA occupancy between experimental states, we expect the peak to have the same proportional share of DNA binding across the experimental states, on average. Likewise, any systematic difference in the DNA binding fold changes for different average DNA binding levels will be the result of biases in the data rather than a genuine biological signal. As such, if there is an equal amount of DNA occupancy between the experimental states, then comparing the proportional share of DNA binding for a peak across the experimental states would correctly normalize the data between sample groups. 

\subsubsection{Normalization by Distribution}
\label{distribution-norm-methods}

Normalization by Distribution methods leverage an aspect of the raw read count distribution (or a function of it) to normalize between sample groups. 

\paragraph{Technical Conditions}
\label{subsec:distribution-technical-conds}
\begin{itemize}
    \item Balanced Differential DNA Occupancy: there is roughly balanced differential DNA occupancy across the experimental states. That is, the number of peaks with more DNA occupancy (per cell) in one experimental state is approximately equal to the number of peaks with more DNA occupancy (per cell) in the other experimental state(s) (see Figure~\ref{fig:balance}).
    \item Peaks with Differential DNA Occupancy and Peaks without Differential DNA Occupancy Behave the Same: the effects of any experimental artifacts on the peaks with differential DNA occupancy are the same as the effects of the experimental artifacts on the peaks without differential DNA occupancy.
\end{itemize}

\begin{figure}[H]
    \centering
    \includegraphics[width=\linewidth, alt = {}]{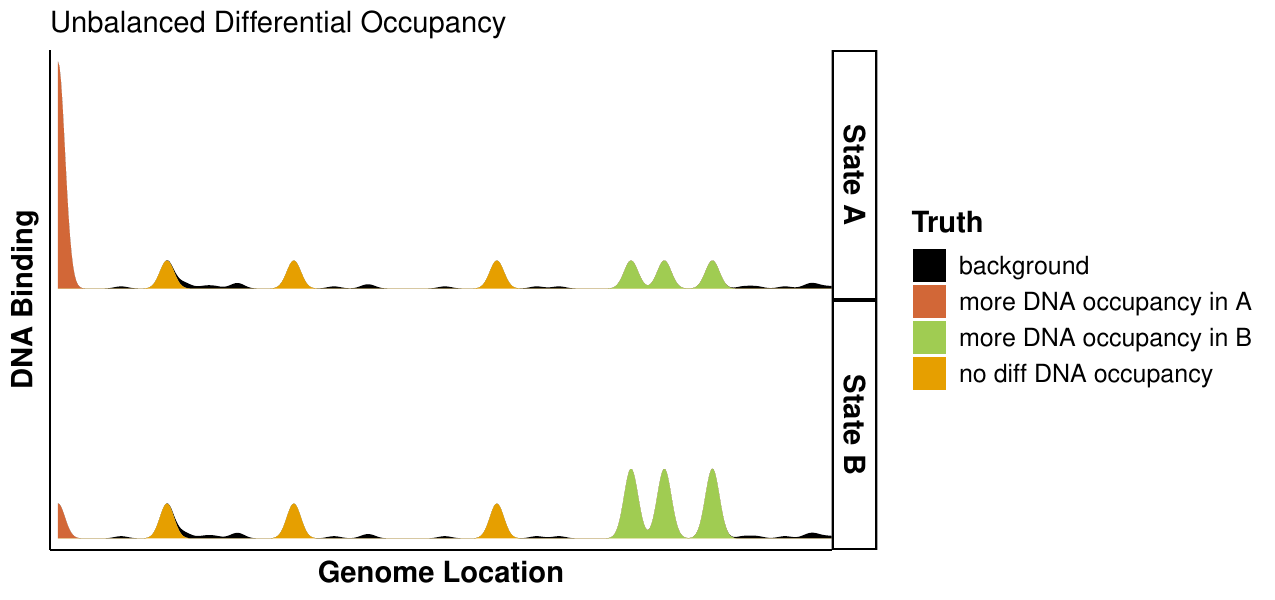}
    \caption{A toy example showing equal total DNA occupancy and constant background but {\bf balanced} differential DNA occupancy across the experimental states, with state A having one peak with more DNA occupancy, and state B having three peaks with more DNA occupancy.}
    \label{fig:balance}
\end{figure}

\paragraph{Methods}
\begin{enumerate}
\item Trimmed Mean of the M-values (TMM) (Reads in Peaks) normalization first selects a sample to serve as the basis of comparison for the other samples.\footnote{TMM selects the sample whose upper quartile of DNA binding is closest to the mean upper quartile of DNA binding as the reference sample by default \cite{EdgeR}.} Then, the M and A values of the signal magnitudes are calculated for each peak in the consensus peakset across the experimental states, relative to the selected reference sample \citep{TMM}.
        
Let $k_{ij}$ be the raw read count associated with consensus peak $i$ in sample $j$, $N_j$ be the total number of reads associated with the consensus peakset in sample $j$ (i.e., $ N_j = \sum_{i} k_{ij}$), and $r$ be the selected reference sample. Then, the M value for consensus peak $i$ in sample $j$, relative to reference $r$, is defined as follows:

\begin{equation}
M_{ij}^r = \frac{\log_2({k_{ij}/N_j})}{\log_2(k_{ir}/N_r)}
\end{equation}
        
Meanwhile, the A value for peak $i$ in sample $j$, relative to reference $r$ is defined as follows:

\begin{equation}
A_{ij}^r = \frac{1}{2}\log_2\biggl(\frac{k_{ij}}{N_j}\cdot\frac{k_{ir}}{N_r}\biggl)
\end{equation}

The $M_{ij}^r$ and $A_{ij}^r$ values are then trimmed twice such that only peaks with central $M_{ij}^r$ and $A_{ij}^r$ values remain \citep{TMM}.\footnote{By default, the $M_{ij}^r$ values are trimmed by $30\%$, and the $A_{ij}^r$ values are trimmed by $5\%$ \cite{TMM}.} The goal of trimming the $M_{ij}^r$ and $A_{ij}^r$ values is to approximate a set of consensus peaks that do not have differential DNA occupancy between the experimental states. However, for trimming the non-central $M_{ij}^r$ and $A_{ij}^r$ values to correctly approximate a set of peaks without differential DNA occupancy, there must be a balanced amount of differential DNA occupancy between the experimental states. Let the set of peaks remaining after trimming the $M_{ij}^r$ and $A_{ij}^r$ values be denoted as $Q$. The set $Q$ is used to calculate the sample-wide size factor for sample $j$ ($s_j$), scaled by the reference sample $r$:
\begin{equation}
    \log_2(s_j) =\frac{\sum_{i \in Q}w_{ij}^rM_{ij}^r}{\sum_{i \in Q}w_{ij}^r}
\end{equation}

\textit{where,} 

    \begin{equation}
        w_{ij}^r = \frac{N_j - k_{ij}}{N_jk_{ij}} + \frac{N_r - k_{ir}}{N_rk_{ir}}
    \end{equation}

Notably, TMM (Reads in Peaks) calculates the sample-wide size factor for sample $j$ only using the subset of consensus peaks in $Q$, i.e., the set which approximates the consensus peaks without differential DNA occupancy between experimental states. As such, TMM (Reads in Peaks) relies on the technical condition that the peaks with differential DNA occupancy behave the same as the peaks without differential DNA occupancy. Otherwise, the effects of experimental artifacts on the peaks without differential DNA occupancy would not be representative of the effects of experimental artifacts on the peaks with differential DNA occupancy.
    
\item Relative Log Expression (RLE) (Reads in Peaks) normalization first finds the ratio of the raw read count associated with consensus peak $i$ in a given sample to the pseudo-reference sample, which is the geometric mean of the raw read count associated with consensus peak $i$ across all of the samples \cite{RLE}. The process of calculating the ratio between the raw read count and the geometric mean of the raw read counts is repeated until a ratio has been generated for every peak in the consensus peakset across experimental states. From here, the median ratio is computed for each sample, and this median ratio serves as the sample's sample-wide size factor \citep{RLE}. That is, the sample-wide size factor for some sample $j$ ($s_j$) using RLE (Reads in Peaks) normalization is defined as follows:

    \begin{equation}
        s_j = \textrm{median}_i\bigg(\frac{k_{ij}}{(\prod_{l =1}^{m} k_{il})^{1/m}}\bigg)
    \label{eq:sizefactor_RLE}
    \end{equation}

\noindent \textit{where $k_{ij}$ is the raw read count associated with consensus peak $i$ in sample $j$ and $m$ is the total number of samples across all the experimental states.} \newline

The purpose of using the median ratio as the sample-wide size factor is to approximate the read count for a peak that does not have differential DNA occupancy between the experimental states \citep{RLE}. For peaks without differential DNA occupancy between experimental states, any difference in the median ratio between would be due to experimental artifacts rather than a genuine biological signal, thereby making the median ratio a meaningful sample-wide size factor. As such, RLE (Reads in Peaks) relies on the median peak not having differential DNA occupancy between the experimental states to correctly normalize the data between samples. Indeed, when there is unbalanced DNA occupancy and a high proportion of peaks with differential DNA occupancy, the peak corresponding to the median ratio would exhibit differential DNA occupancy between experimental states, making the median ratio a poor basis of normalization. Moreover, it follows that RLE (Reads in Peaks) also has the technical condition that the peaks without differential DNA occupancy behave the same as the peaks with differential DNA occupancy given that RLE (Reads in Peaks) generates each sample's size factor by approximating a ratio for a consensus peak without differential DNA occupancy between experimental states. 

\item MAnorm2 is a normalization method specifically designed for ChIP-seq data by Tu et al.\ \cite{MAnorm2}. MAnorm2 first normalizes the samples within a given experimental state. Then, the data is normalized across experimental states. Both of MAnorm2's normalization steps utilize the same general algorithm, which we describe in the subsequent paragraphs. 

Let $k_{ij}$ denote the raw read count associated with peak $i$ in sample $j$. In MAnorm2, the normalized read count for peak $i$ in sample $j$ relative to sample $l$, which we denote as $k_{ij}^*$, is defined as follows:

\begin{equation}
    k_{ij}^* = \alpha_j + \beta_j \cdot \log_2(k_{ij})
\label{eq:MAnorm_linear_trans}
\end{equation}

Notice that MAnorm2 does not generate a sample-wide size factor for sample $j$. Instead, each peak's normalized read count is a linear function of its log-2 transformed raw read count. The linear coefficients $\alpha_j$ and $\beta_j$ associated with sample $j$ in Equation~\eqref{eq:MAnorm_linear_trans} are computed by comparing the raw read counts in the common peaks across the two samples. A peak is considered a \textit{common peak} if it is called as a peak within both samples \cite{MAnorm2}. The coefficients $\alpha_j$ and $\beta_j$ in Equation~\eqref{eq:MAnorm_linear_trans} are defined as follows, where $[k_{j}]$ is a vector with entries consisting of the raw read count associated with common peak $i$ in sample $j$, and $sd_i$ denotes the sample estimate of the standard deviation in the raw read counts across the common peaks $i$ within sample $j$:

\begin{equation}
    \alpha_j = \textrm{mean}_i\biggl(\log_2([k_{l}])) - \beta_j \cdot \textrm{mean}_i(\log_2([k_{j}]))\biggl)
\label{eq:alpha}
\end{equation}

\begin{equation}
    \beta_j = \frac{sd_i(\log_2([k_{l}]))}{sd_i(\log_2([k_{j}]))}
\label{eq:beta}
\end{equation}

Let $[M^*]$ denote the vector of the normalized M values in the common peak regions across samples $j$ and $l$, and $[A^*]$ denote a vector of the normalized A values in the common peak regions across samples $j$ and $l$. Then, $[M^*]$ and $[A^*]$ are computed as follows, where $[k_j^*]$ is a vector consisting of the normalized read counts in sample $j$ associated with each peak $i$ (i.e., $k_{ij}^*$), and $k_{l}$ is a vector consisting of the raw read counts associated with sample $l$ for each peak $i$:

\begin{equation}
    [M^*] = [k_{j}^*] -  \log_2([k_{l}])
\label{eq:m-val}
\end{equation}

\begin{equation}
    [A^*] = \frac{1}{2}\biggl(\log_2([k_{l}]) + [k_{j}^*] \biggl)
\label{eq:a-val}
\end{equation}

Based on the definitions of $\alpha_j$ and $\beta_j$, the $[M^*]$ and $[A^*]$ vectors associated with the two samples will satisfy the following two conditions under MAnorm2 normalization:

\begin{equation}
\textrm{mean}_i([M^*]) = 0 
\label{eq:MAnorm_constraint_1}
\end{equation}

\begin{equation}
    \textrm{cov}_i([M^*], [A^*]) = 0
\label{eq:MAnorm_constraint_2}
\end{equation}

Equations~\eqref{eq:MAnorm_constraint_1} and \eqref{eq:MAnorm_constraint_2} guarantee that the ordinary least squares line on the normalized MA plot is the horizontal axis. When there is unbalanced differential occupancy between experimental states, the ordinary least squares line on the raw MA plot is skewed towards the experimental state with more differential DNA occupancy per cell, which is to say that the raw M-values will be more extreme for the states with higher DNA occupancy. In turn, when the read counts are normalized such that Equations~\eqref{eq:MAnorm_constraint_1} and \eqref{eq:MAnorm_constraint_2} are satisfied, the normalized M values associated with peaks that do not have differential DNA occupancy between states will be pulled away from zero (on the log-2 scale) to offset the unbalanced distribution of the M values, thereby leading to false discoveries of differentially bound peaks. Therefore, MAnorm2 relies on the technical condition that there is balanced differential DNA occupancy between experimental states.\footnote{In the original paper introducing MAnorm2, Tu et al.\ note that MAnorm2 assumes that there is ``no global change of protein binding intensities" in the common peak regions \cite{MAnorm2}. From our understanding of the method's details, they are describing balanced differential occupancy rather than an equal total amount of DNA occupancy in this statement.}
\end{enumerate}

\paragraph{Motivation for Distribution Methods}
Normalization by Distribution methods work by supposing that we can normalize the raw read counts by comparing characteristics of the read count distribution between samples. If most peaks do not have differential DNA occupancy between the experimental states, then the central measurements of the raw read count distribution will correspond to a peak without differential DNA occupancy between states. Additionally, if the effects of experimental artifacts on the peaks that have differential DNA occupancy are the same as the effects of experimental artifacts on the peaks that do not have differential DNA occupancy, then we can correctly normalize the entire sample by using the sample-wide size factor that we computed by approximating a peak \textit{without} differential DNA occupancy between the experimental states.

\subsubsection{Normalization by Background}
\label{background-norm-methods}
Normalization by Background methods first partition the chromosomes into background bins that are roughly 15,000 base pairs long \citep{DiffBind-vignette}. For Normalization by Background methods, the background bins rather than the consensus peakset serve as the normalization unit (see Figure~\ref{fig:bin_obs_unit}).

\paragraph{Technical Conditions}
\label{subsec:background-technical-conds}
\begin{itemize}
    \item Equal Amount of Background Binding: the number and distribution of rogue reads (i.e., binding in regions that are not truly occupied by the protein of interest) within each sample is roughly the same across the experimental states.
    \item Previous Technical Conditions: the aforementioned technical conditions for any given method must be met by the background bins across the experimental states. Because the background has zero differential occupancy (by definition), when using bins as normalization units, the other technical conditions must also be met when describing the DNA binding across the background bins.
\end{itemize}

\paragraph{Methods}

\begin{enumerate}
    \item Library Size (Background Bins) was previously described in the Normalization by Library Size methods section. To summarize: a sample is normalized using Library Size (Background Bins) by dividing the raw read count associated with each peak by the total raw read count summed over the background bins \citep{TotalCount}.

    \item Trimmed Mean of the M-Values (TMM) (Background Bins) works very similarly to TMM (Reads in Peaks). The key difference is that the unit used to generate the sample-wide size factor is the background bins rather than the consensus peakset across experimental states for TMM (Background Bins).
    
    \item Relative Log Expression (RLE) (Background Bins) uses the same normalization procedure as RLE (Reads in Peaks), but, like with  TMM (Background Bins), the background bins are used to generate the sample-wide size factor rather than the consensus peakset across experimental states for RLE (Background Bins).
\end{enumerate}

\paragraph{Motivation for Background Methods}
Peaks are typically only a few hundred base pairs long. So, any substantial (non-trivial) difference in the amount of DNA binding in a given background bin across experimental states is expected to be the result of different experimental artifacts between samples rather than true biological differences in the amount of DNA occupancy between experimental states \cite{DiffBind-vignette}. Thus, we can leverage background bin methods to correctly normalize between samples, provided that the technical condition of equal background binding across experimental states is satisfied. 

\subsubsection{Normalization by Spike-in DNA}
\label{control-norm-methods}

When experimental data violates the technical conditions listed in the previous sections, Normalization by Spike-in DNA can enable us to correctly normalize between ChIP-seq samples. Normalization by Spike-in DNA works by amending the ChIP-seq data collection procedure by adding a fixed amount of DNA from an organism that is not interrogated in the experiment prior to immunoprecipitation. For example, in a ChIP-seq experiment using mammalian cells, yeast DNA could be added. The assumption is that any yeast DNA that remains in the final sample represents background, since none of the yeast DNA would be occupied by the mammalian protein of interest. Figure~\ref{fig:spike-in-illustration} provides a toy example that compares the experimental ChIP-seq data with the data generated from spike-in controls.

\begin{figure}[H]
    \centering
    \includegraphics[width=\linewidth, alt = {}]{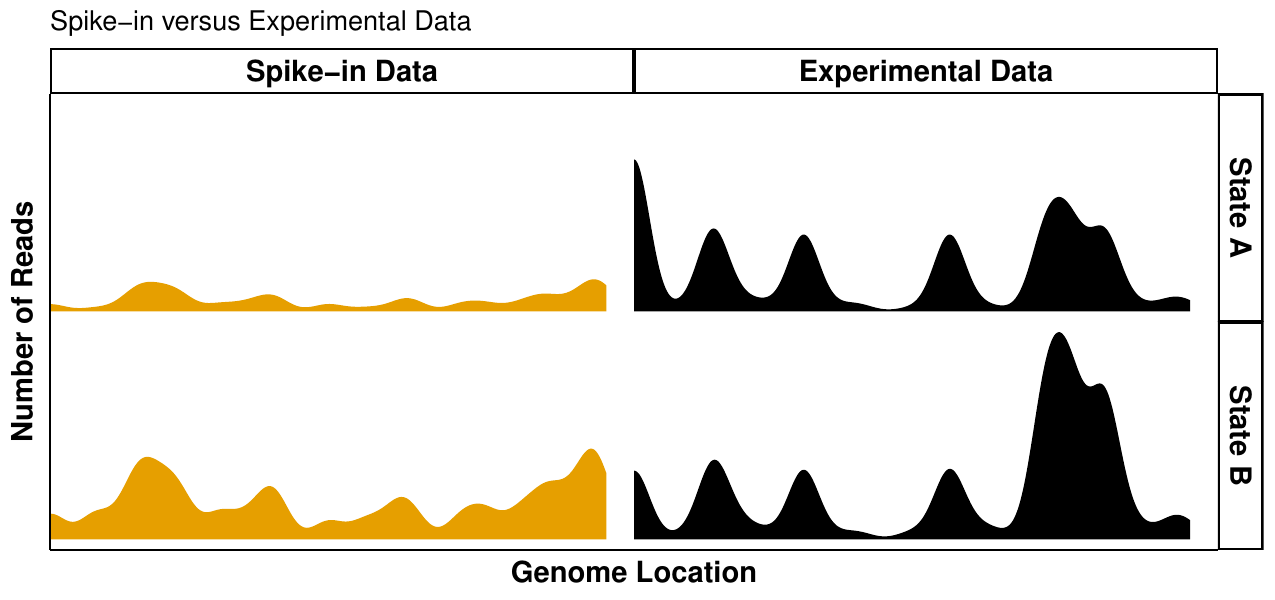}
    \caption{Spike-in Control Illustration. The horizontal axis is the genome location, and the vertical axis is the number of reads aligned to that genome location. The plot is faceted by spike-in versus experimental data as well as by experimental state (A versus B). The left side of the plot (in yellow) corresponds to the spike-in DNA data. Note how there are no sharp peaks in the spike-in data since there should be no changes in DNA occupancy within the spike-in control samples. In contrast, there are sharp peaks on the right side of the plot (in black), which correspond to the experimental data. Note that there are more reads in the spike-in control data for state B than state A (compare the right plots). Normalization with spike-in DNA would aim to eliminate the discrepancy between the spike-in sample read counts across the two states.}
    \label{fig:spike-in-illustration}
\end{figure}

\paragraph{Technical Conditions}
\begin{itemize}
    \item Existence of Spike-in Controls: the spike-in DNA was added by the experimenter, and it comes down in the immunoprecipitation with the same frequency regardless of the experimental state.
    \item Spike-in Controls Behave like Samples of Interest: the effects of experimental artifacts on the spike-in DNA reflect the effects of experimental artifacts on the experimental samples.
\end{itemize}

\paragraph{Method}

\begin{enumerate}
\item Normalization by Spike-in DNA works by adding spike-in DNA prior to immunoprecipitation \cite{Spikein}. If the same amount of spike-in DNA is added to each sample, then the total number of reads is expected to be the same across the samples, too \cite{spike-in-data-collection}. Thus, any difference in the total number of reads across spike-in control samples is taken to be the result of differences in the experimental artifacts between the samples. So, sample-wide size factors for each sample can be calculated by equalizing the spike-in signal between samples \cite{spike-in-data-collection}. Since the spike-in DNA is not expected to have any differential DNA occupancy between states, a Normalization by Background method is used to calculate sample-wide size factors for each sample \cite{DiffBind-vignette}. The sample-wide size factors calculated using the spike-in DNA are then applied to the relevant experimental sample in order to ameliorate the impact of experimental artifacts on the read count differences between experimental samples. 
\end{enumerate}

\paragraph{Motivation for using Spike-in DNA}

We often have an \textit{a priori} understanding that the amount of spike-in DNA that comes down in immunoprecipitation should remain the same across experimental states \citep{Spikein}. If the amount of spike-in DNA at a particular genomic location misaligns with this expectation, we can take this misalignment to be the result of differences in the experimental artifacts between the samples rather than genuine differences in the biological signal. Moreover, assuming that the impact of experimental artifacts in the spike-in control samples reflects the impact of experimental artifacts in the experimental samples of interest, we can apply the sample-wide size factors calculated from the spike-in DNA to the relevant experimental sample to correctly normalize the experimental data between samples.

\subsubsection{The Oracle Normalization Method}
\label{subsubsec:oracle}

In our simulation results, an important basis of comparison is the omniscient Oracle normalization method, which computes the correct sample-wide size factor based on the population-level simulation parameters. Specifically, the Oracle uses the parameter values that calculate the expected value of the negative binomial distribution, which represents the expected amount of DNA binding within a given peak. Let $p_{ij}$ denote the proportional share of DNA occupancy in peak $i$ for sample $j$ before we simulate any differential DNA occupancy between states, and $(fc)_{ij}$ represent the fold change in DNA occupancy for peak $i$ in sample $j$ relative to a sample in the other experimental state. Finally, let $\textrm{(library multiplier)}_j$ denote the scalar used to increase the variability in DNA binding between samples. Then, the expected value of the negative binomial distribution, which represents the mean amount of DNA binding in the peak, is defined as follows:

\begin{equation}
    \mu_{ij} = \frac{p_{ij} \cdot (fc)_{ij}}{(\textrm{basesum})_j} \cdot (\textrm{library multiplier})_j \cdot 600,000
    \label{eq:simulated_mu}
\end{equation}

\textit{where,}

\begin{equation}
(\textrm{basesum})_j = [(fc)_j]^\intercal \cdot [p_j]   
\label{eq:basesums}
\end{equation}

\textit{and 600,000 is the expected library size for all samples.}

Based on how we compute the expected value of the negative binomial distribution, we define the Oracle's size factor for sample $j$ as follows, where $m$ is the total number of samples across the two experimental states we simulate:

\begin{equation}
    s_j = \frac{(\textrm{normFactor})_j}{\prod_{h = 1}^{m}{(\textrm{normFactor})_h}^{(1/m)}}
\label{eq:sizefactor_oracle}
\end{equation}

\textit{where, }

\begin{equation}
    \textrm{(normFactor})_j = \frac{(\textrm{library multiplier})_j}{(\textrm{basesum})_j}
\label{eq:normfactor}
\end{equation}

As a result, we obtain the following when we divide the mean read count for peak $i$ under sample $j$ (i.e., $\mu_{ij}$) by the normalization factor for sample $j$:

\begin{equation}
    \frac{\mu_{ij}}{\textrm{(normFactor)}_j} = p_{ij} \cdot (fc)_{ij} \cdot 600,000
\label{eq:norm_read_counts}
\end{equation}

Since $p_{ij}$ (i.e., the base proportion of DNA occupancy in peak $i$ in sample $j$) and the expected library size of $600,000$ are constant across all the samples in our simulation, only $(fc)_{ij}$ changes between experimental states. Therefore, when we compare read counts that are normalized by the Oracle across experimental states, we correctly estimate the fold change in DNA occupancy across these states. The geometric mean of normalization factors in the denominator of Equation \eqref{eq:sizefactor_oracle} serves as the pseudo-reference sample for the normalization. In the next section, we provide a toy example to walk through the Oracle and the population-level simulation parameters it utilizes to compute the sample-wide size factor for each replicate.

\paragraph{Oracle Toy Example}

Suppose there are two experimental states (A and B) and three peaks within each experimental state (denoted as Peaks 1,2, and 3). The library multipliers for state A and state B are $0.9$ and $0.8$, respectively. We can find the Oracle normalization factors associated with states A and B in our toy example in Figure \ref{fig:Oracle-illustration} by using Equation \eqref{eq:normfactor} and leveraging the fact that the library multipliers for state A and state B are $0.9$ and $0.8$, respectively. Namely, the normalization factors associated with A and B are the following:

\begin{equation}
\textrm{(normFactor})_A = \frac{0.9}{1} = 0.9
\label{eq:normfactor_ex_A}
\end{equation}

\begin{equation}
\textrm{(normFactor})_B = \frac{0.8}{\frac{4}{3}} = 0.6
\label{eq:normfactor_ex_B}
\end{equation}

Then, we can find the Oracle size factor for each experimental state by using Equation \eqref{eq:sizefactor_oracle}.

\begin{equation}
s_A = \frac{0.9}{\sqrt{0.6 \cdot 0.9}} = 1.2247
\label{eq:sizefactor_ex_A}
\end{equation}

\begin{equation}
s_B = \frac{0.6}{\sqrt{0.6 \cdot 0.9}} = 0.8165
\label{eq:sizefactor_ex_B}
\end{equation}

Dividing the raw read counts for states A and B given in Figure \ref{fig:Oracle-illustration}(d) by their respective Oracle size factors of $1.04447$ and $0.95743$ gives us the Oracle normalized read counts given in Figure \ref{fig:Oracle-illustration}(e). Table \ref{tab:oracle_read_calcs} shows the calculations for the Oracle-normalized read counts. 

\begin{figure}[H]
    \centering
    \includegraphics[width = \linewidth, alt = {}]{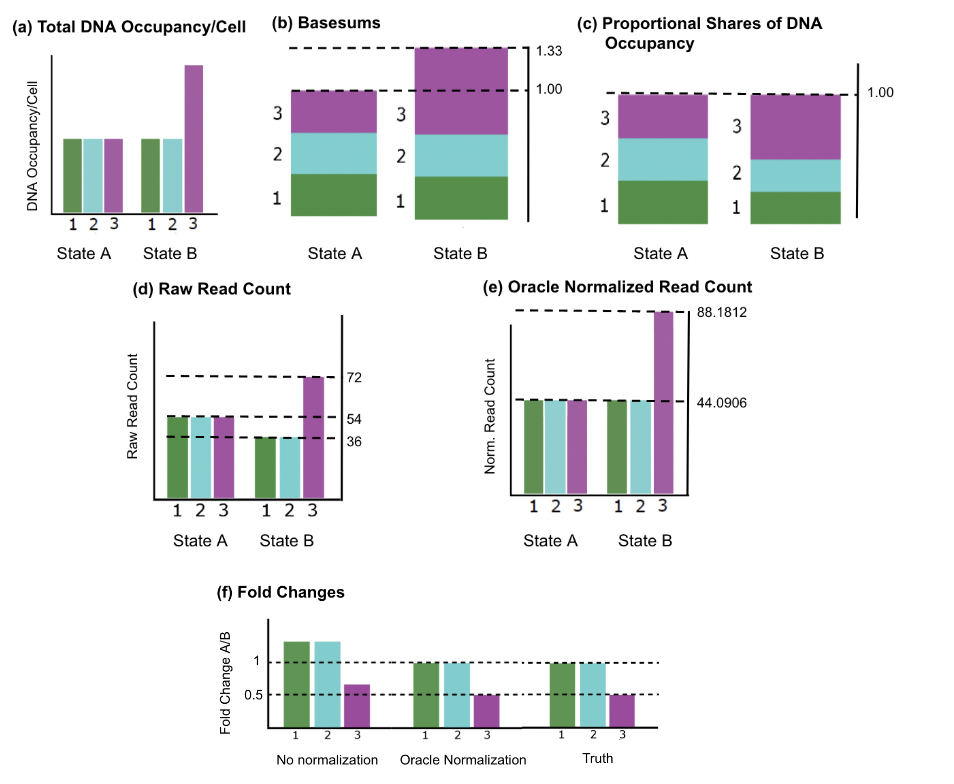}
    \caption[Oracle Normalization Method Illustration]{Illustration of Oracle Normalization Method. \textbf{(a)} represents the DNA occupancy (per cell) for each peak in states A and B.  \textbf{(b)} is the corresponding basesums for states A and B.  \textbf{(c)} provides the proportional shares of DNA occupancy for the peaks for states A and B. The raw read counts are generated through the general process outlined in Equation \eqref{eq:simulated_mu} and are depicted in \textbf{(d)}. Using the raw read counts, Peaks 1 and 2 appear differentially bound even though they do not have different DNA occupancy in \textbf{(f)}. Additionally, even though Peak 3 has 1/2 as much DNA occupancy in state A as compared to B in \textbf{(a)}, it appears to have 2/3 as much in \textbf{(f)}. Note that when we normalize our toy data using the Oracle, the fold changes associated with the Oracle normalized read counts in \textbf{(e)} match the \textit{true} fold changes in the DNA occupancy between the states depicted in \textbf{(f)}.}
    \label{fig:Oracle-illustration}
\end{figure}

From the Oracle-normalized read counts computed in Table \ref{tab:oracle_read_calcs}, we would conclude that Peaks 1 and 2 are \textit{not} differentially bound and that Peak 3 \textit{is} differentially bound, with two times more DNA binding in state B than state A. As shown in Figure ~ref{fig:Oracle-illustration}~\textbf{(e)} and \textbf{(f)}, the Oracle-normalized DNA binding results correctly align with the amount of DNA occupancy (per cell) depicted in Figure~\ref{fig:Oracle-illustration}~\textbf{(a)}. Thus, the toy example demonstrates that the Oracle normalization method correctly normalizes between samples by leveraging the population-wide simulation parameters.

\begin{table}[H]
\caption{Oracle Normalized Read Counts. The columns correspond to the experimental state associated with the raw read count, and the rows correspond to the peak associated with the raw read count. Each entry is the Oracle normalized read count for the specific peak, which is found by multiplying the raw read count depicted in  Figure~\ref{fig:Oracle-illustration}~\textbf{(d)} by the size factors calculated in Equations \eqref{eq:sizefactor_ex_A} and \eqref{eq:sizefactor_ex_B}.
\label{tab:oracle_read_calcs}}%
\begin{tabular*}{\columnwidth}{@{\extracolsep\fill}llll@{\extracolsep\fill}}
\toprule
& Norm. read count (A) & Norm. read count  (B) \\
\hline
Peak 1 & $54/1.2247 = 44.0906$ & $36/0.8165 = 44.0906$\\
Peak 2 & $54/1.2247 = 44.0906$ & $36/0.8165 = 44.0906$\\
Peak 3 & $54/1.2247 = 44.0906$ &  $72/0.8165 = 88.1812$\\
\end{tabular*}
\end{table}

\subsection{Simulation Results: 2 Replicates per Experimental State}
\label{appendix-10-rep-results}

Figures \ref{all_FDR_2_rep} and \ref{all_power_2_rep} provide the simulation results with two replicates per experimental state. The overall trends observed in the average empirical FDR and power for two replicates per experimental state are consistent with the results for three replicates per experimental state (see Figures \ref{fig:FDR-all} and \ref{fig:Power-all}). With fewer replicates (two instead of three), we expect fewer overall discoveries. Thus, it is not surprising that both the FDR and the power decrease when we have two replicates per experimental state instead of three.

\newpage

\begin{figure}[H]
\centering
\includegraphics[width = \linewidth, alt = {}]{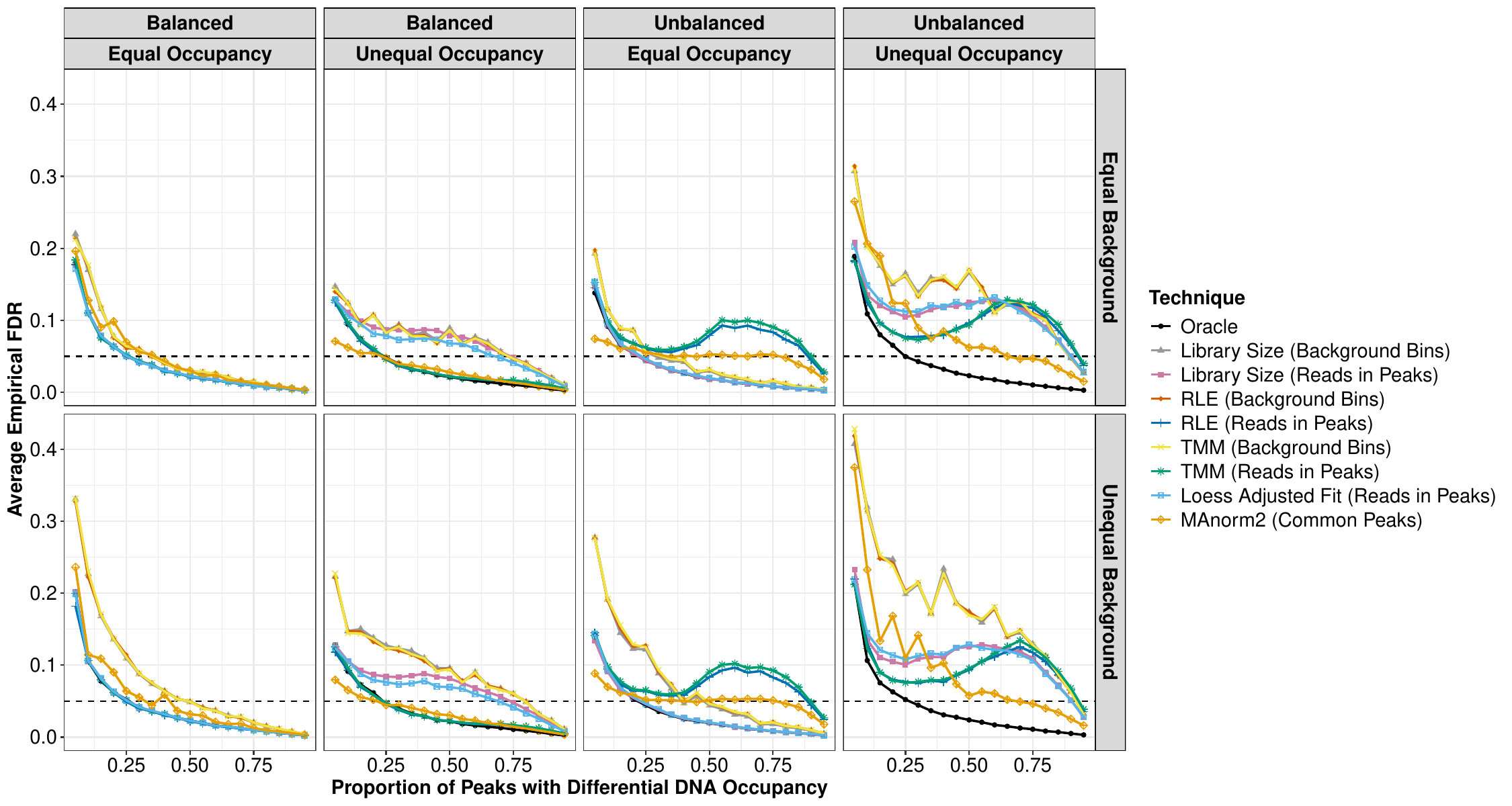}
\caption{The average empirical false discovery rate (FDR) for each normalization method with two replicates per experimental state, faceted by simulation condition. The horizontal axis is the proportion of peaks with differential DNA occupancy, and the vertical axis is the average empirical false discovery rate (FDR). When a curve is below the dashed horizontal black line, then the associated normalization method (on average) successfully controls the false discovery rate at a level of $0.05$ for the given proportion of peaks with differential DNA occupancy. We consider normalization methods that track with the Oracle (the solid black line) to be performing well with respect to the average empirical false discovery rate (FDR) in our simulation.}
\label{all_FDR_2_rep}
\end{figure}

\begin{figure}[H]%
\centering
\includegraphics[width = \linewidth, alt = {}]{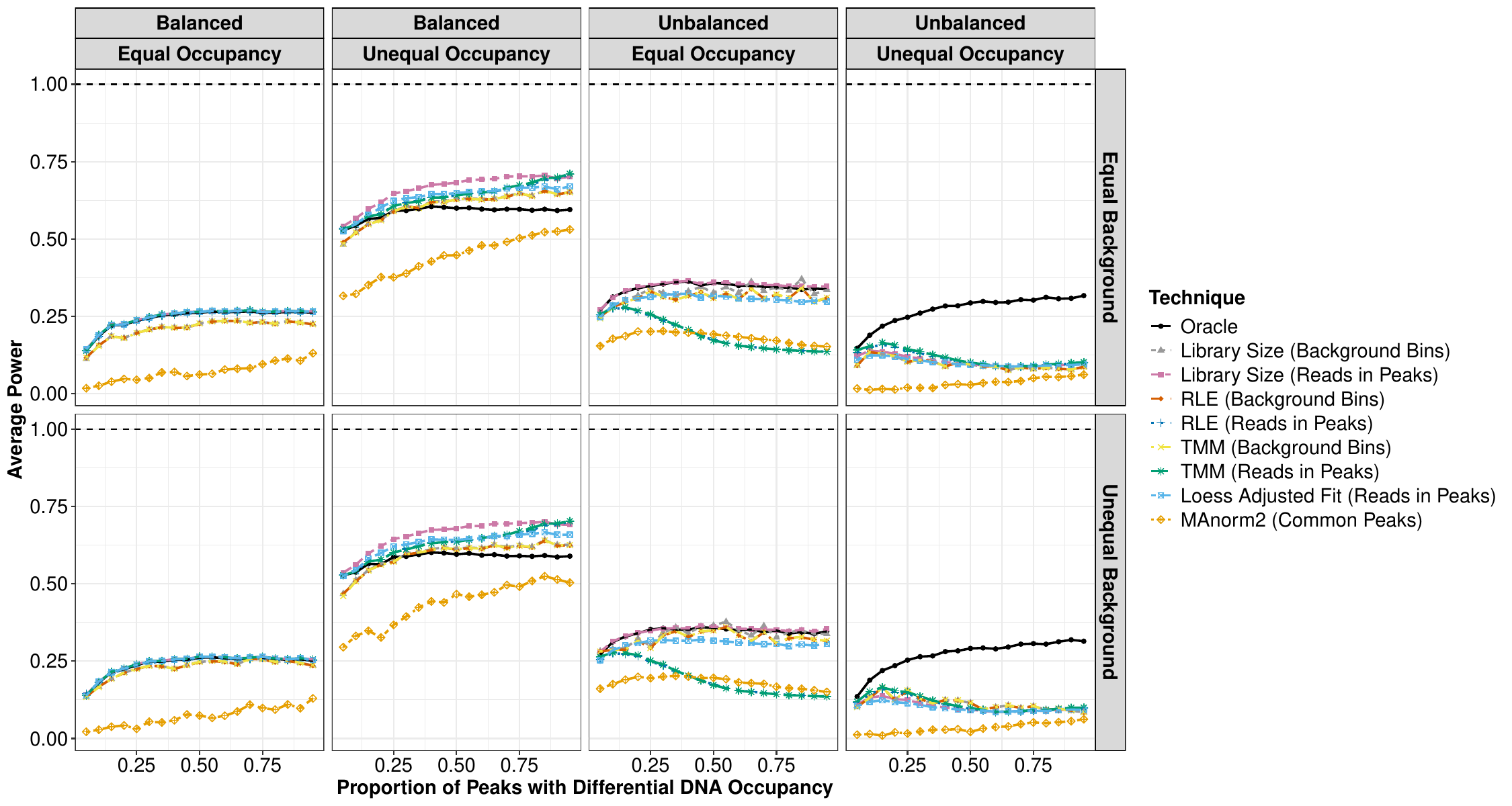}
\caption{The average power for each normalization method with two replicates per experimental state, faceted by simulation condition. The horizontal axis is the proportion of peaks with differential DNA occupancy, and the vertical axis is the average power. The nearer a curve is to the dashed horizontal black line at $1$, the higher power the associated normalization method has (on average) for the given proportion of peaks with differential DNA occupancy.}
\label{all_power_2_rep}
\end{figure}

\end{document}